\documentclass[journal,comsoc,10pt]{IEEEtran}

\usepackage[T1]{fontenc}

\usepackage{makecell}
\usepackage{booktabs}
\usepackage{amssymb}
\usepackage{threeparttable}

\usepackage{makecell}
\usepackage{cite}
\usepackage[colorlinks,
linkcolor=red,
anchorcolor=blue,
citecolor=green
]{hyperref}

\ifCLASSINFOpdf
\else
\fi
\usepackage{amsmath}
\usepackage{diagbox}

\usepackage{ amssymb }
\usepackage{amsfonts}
\usepackage{caption}
\usepackage{graphicx}
\usepackage{lettrine}
\usepackage{epstopdf}
\usepackage{textcomp,booktabs}
\usepackage[usenames,dvipsnames]{color}
\usepackage{colortbl}
\definecolor{mygray}{gray}{.9}
\definecolor{mypink}{rgb}{.99,.91,.95}
\definecolor{mycyan}{cmyk}{.3,0,0,0}
\usepackage{pifont}
\usepackage{subfigure}
\usepackage{multirow}
\usepackage{url}
\usepackage{color}
\usepackage{threeparttable}
\usepackage{setspace}
\usepackage{lineno}
\usepackage{tabularx}
\usepackage[dvipsnames,usenames]{color}

\usepackage{bm}

\usepackage{algorithm} 
\usepackage{algorithmic} 

\usepackage{dblfloatfix}

\usepackage[dvipsnames,usenames]{color}
\usepackage[usenames,dvipsnames]{color}

\definecolor{light-gray}{gray}{0.90}
\begin{document}

	\title{Intention-Aware Semantic Agent Communications for AI Glasses}

	\author{Peiwen Jiang,~\IEEEmembership{Member,~IEEE,} Fangyu Liu, ~\IEEEmembership{Graduate Student Member,~IEEE,} Jiajia Guo,~\IEEEmembership{Member,~IEEE,}  Chao-Kai Wen,~\IEEEmembership{Fellow,~IEEE,}  Shi Jin,~\IEEEmembership{Fellow,~IEEE,} and Jun Zhang,~\IEEEmembership{Fellow,~IEEE}
			\thanks{P. Jiang, J. Guo and J. Zhang are with the Department of Electronic and Computer Engineering,
				Hong Kong University of Science and Technology, Hong Kong (e-mail: eepwjiang@ust.hk; eejiajiaguo@ust.hk;  eejzhang@ust.hk). (Corresponding author: Jun Zhang)}
			\thanks{C.-K. Wen is with the Institute of Communications Engineering, National
				Sun Yat-sen University, Kaohsiung 80424, Taiwan (e-mail: chaokai.wen@mail.nsysu.edu.tw).}
                \thanks{F. Liu and S. Jin are with the School
				of Information Science and Engineering, Southeast University, Nanjing
				210096, China (e-mail:  fangyuliu@seu.edu.cn; jinshi@seu.edu.cn).}}
	
	\maketitle
	\pagestyle{empty}  
	\thispagestyle{empty} 

\begin{abstract}
Smart glasses are emerging as a promising interface between humans and artificial intelligence (AI)  agents, enabling first-person perception, contextual awareness, and real-time assistance. However, continuous offloading of visual data from wearable devices to cloud-based vision-language models (VLMs) is fundamentally constrained by limited wireless bandwidth and energy resources. This paper proposes an intention-aware semantic agent communication framework for AI glasses, where data transmission is guided by user intention rather than raw pixel fidelity. In the proposed architecture, AI glasses act as an edge semantic agent while a server-side VLM executes high-level cognition and reasoning.  The user intention can be inferred by the server-side VLM through the current transmitted content and the historical prompts. Driven by specific user intentions, the glasses adaptively preserve textual content, document layout, or object semantics before transmission. We evaluate three representative scenarios with different lightweight preprocessing tools on the AI glasses. Simulation results demonstrate that intention-aware preprocessing significantly achieves more than 50\% bandwidth reduction depending on the current task while maintaining task performance. Moreover, semantic transmission exhibits graceful degradation under low SNRs. The findings demonstrate that aligning communication resources with user intention is essential for robust and efficient wearable AI agent systems.

\end{abstract}


\section{Introduction}

\IEEEPARstart{A}{rtificial} intelligence (AI) glasses are emerging as a lightweight and ubiquitous interface between humans and AI agents, enabling continuous first-person visual perception and contextual assistance \cite{waisberg2024meta,wang2025systematic,chen2023applying}. By coupling wearable sensing with cloud-based large vision-language models (VLMs), AI glasses can support real-time document understanding, scene reasoning, and task guidance. However, their compact form factor imposes stringent constraints on the computational capability, battery capacity, and wireless bandwidth \cite{chen2024enabling, bariah2024ai}. As a result, most high-level reasoning must be offloaded to edge or cloud servers via wireless links \cite{ xu2024unleashing, shen2023large}. Within such wearable AI systems, communication is no longer a background utility but a critical resource that directly determines latency, reliability, and user experience. The fundamental challenge is therefore how to efficiently transmit task-relevant visual information under severe wireless communication constraints.

Conventional wireless communication systems are designed to reliably transmit bits with minimal distortion, following the Shannon paradigm that treats all bits equally important \cite{bao2011towards}. In this framework, performance is evaluated through bit error rate, spectral efficiency, or distortion metrics such as PSNR and SSIM, without explicit awareness of downstream tasks. However, in wearable AI scenarios, the objective is not faithful pixel reconstruction but accurate task completion at the receiver. For instance, answering a question about a document, extracting key entities, or describing a scene does not require lossless image recovery, but rather the preservation of task-relevant semantics. This shift from bit-level fidelity to task effectiveness fundamentally challenges traditional communication design principles. Recent advances in semantic communications advocate transmitting meaning instead of raw data by jointly optimizing source representation, channel coding, and inference objectives \cite{gunduz2022beyond, qin2021semantic, shi2021semantic, yang2022semantic}. Through the application of AI techniques, these semantic transmission systems further demonstrate that aligning communication strategies with task goals can significantly improve transmission efficiency under bandwidth constraints \cite{xie2020deep, wang2022wireless, ma2023task, zhang2024intellicise}.

With the emergence of large foundation models, semantic communication has entered a new paradigm. Instead of designing lightweight, task-specific encoders, recent works exploit pretrained large language models (LLMs), VLMs, and diffusion-based generative models as shared semantic knowledge bases for communication systems \cite{jiang2023large, jiang2025semantic, li2024end}. These models inherently encode rich world knowledge and cross-modal alignment, enabling semantic compression through reasoning, abstraction, and generative reconstruction rather than explicit signal recovery. Generative semantic communication frameworks also take advantage of large models to reconstruct semantically consistent content on the receiver side \cite{grassucci2023generative, li2024end, xu2024semantic}. Meanwhile, the vision of  LLM-empowered wireless intelligence highlights that communication systems are evolving from data pipelines to knowledge-driven infrastructures \cite{shao2024wirelessllm, jiang2024position, ferrag2025llm}. This paradigm suggests that communication should be designed with LLMs that perform reasoning, planning, and decision-making.

Despite these advances, most existing semantic communication frameworks remain task-static, where the semantic coding is optimized for a predefined objective such as image reconstruction, text transmission, video conferencing, or specific task inference \cite{xie2020deep, wang2022wireless, jiang2022deep}. The communication goal is typically fixed during system deployment, assuming a stable downstream task and distortion criterion. However, wearable AI agents operate in highly dynamic and user-driven environments, where communication objectives evolve continuously according to user intentions. The user may alternately request text extraction, structural document reasoning, scene understanding, or goal-directed interaction. In such settings, uniformly transmitting all visual content is inefficient, while designing separate semantic encoders for every possible task remains impractical. This challenge motivates a transition from task-oriented semantic communication to intention-aware  semantic agent communication, where communication policies are adaptively aligned with goals, contextual knowledge, and multi-agent collaboration \cite{zou2023wireless, chen2024enabling, peyrard2024agentic, yu2025semantic,zhou2024semantic}. Emerging studies on semantic agent communications and  multi-agent semantic networking further suggest that communication should become an active component of agent reasoning and planning loops rather than a passive transmission module \cite{chen2025goal, charalambous2025toward}. Therefore, a fundamental challenge arises: \emph{how can communication strategies be dynamically co-optimized with evolving user intentions and agent cognition in wearable AI systems?} However, the current studies cannot dynamically deal with the changing user requirements and adapt to the computation capabilities of the AI glasses.

To address this challenge, we propose an intention-aware semantic agent communication framework for AI glasses. In contrast to conventional  semantic encoders, the proposed framework treats the user intention as a dynamic and latent objective that guides what information should be transmitted. Such intention may originate from historical interaction commands, contextual cues, or autonomous inference by the wearable agent based on the captured visual content. Instead of uniformly transmitting full-resolution images, the edge-side agent selectively extracts and compresses semantically relevant components that are most informative for the inferred intention. The cloud-side agent, built upon powerful VLMs, then performs high-level reasoning. In this architecture, communication becomes an adaptive semantic interface between embodied perception and remote cognition, rather than a passive bit-delivery pipeline. The main contributions of our work are summarized as follows:

\begin{itemize}

\item \textbf{Intention-aware semantic communication pipeline:}  
We propose an intention-aware semantic communication pipeline in which user intention is explicitly modeled as a dynamic control prompt inferred by a cloud-based VLM. Unlike conventional task-specific schemes, the proposed pipeline periodically updates the prompts and tool selections based on user interaction history and visual observations. This intention  prompt determines what semantic features should be extracted using the existing toolbox, allowing adaptive and context-consistent communication strategies.

\item \textbf{Agent-oriented wearable-cloud collaborative system:}
We design a system-level semantic agent architecture for the above pipeline. In this framework, the AI glasses function as an edge perception agent responsible for lightweight visual preprocessing tools, while the cloud-based VLM serves as a cognitive reasoning agent that performs high-level inference and task planning. Communication between the two is structured as an adaptive semantic interface rather than raw data streaming. The cloud agent generates task-specific prompts and tool instructions, and the edge agent executes corresponding lightweight processing and transmission policies under wireless bandwidth constraints. This architecture clearly separates perception, reasoning, and control across edge-cloud boundaries.

\item \textbf{Resolution-robust semantic image transmission method:}
We develop a resolution-robust semantic transmission framework to ensure robustness under varying visual inputs. The processed image may have different resolutions and requires different bandwidth. Specifically, we train a fully convolutional encoder-decoder using multi-scale randomly cropped images, enabling the model to process and reconstruct semantic representations across diverse spatial resolutions without retraining. This design allows the transmission module to flexibly convert the inputted image into the transmission symbols with a fixed compression ratio, while maintaining stable reconstruction quality under bandwidth constraints. 

\end{itemize}

The remainder of this paper is organized as follows. Section \uppercase\expandafter{\romannumeral3} describes the system model and the glass-server communication process. Section \uppercase\expandafter{\romannumeral4} presents the proposed framework. Section \uppercase\expandafter{\romannumeral5} evaluates the performance of the proposed transmission under different glass scenarios, where the transmission overhead is significantly reduced with task performance well preserved. Section \uppercase\expandafter{\romannumeral6} concludes the paper.

\section{Related Works}
In this section, we review research related to semantic agent communication. We first outline semantic and task-oriented communication frameworks, followed by generative and foundation model-based approaches. Then, we discuss edge-cloud multi-agent architectures and wearable AI systems. This overview highlights the need to integrate intention modeling with transmission adaptation for wearable semantic communication.
\subsection{Semantic and Task-Oriented Communication}

Semantic communication departs from the conventional Shannon paradigm by shifting the design objective from bit-level fidelity to meaning-aware and task-oriented performance. Early discussions on semantic information emphasized that reliable symbol transmission does not necessarily guaranty the execution of the task at the receiver \cite{bao2011towards}. Recent advances formalize this shift by introducing semantic entropy, semantic mutual information, and task-driven objectives into communication system design \cite{gunduz2022beyond, shi2021semantic, yang2022semantic, chaccour2024less}. These works establish the theoretical foundation for semantic communication networks that prioritize task effectiveness over raw reconstruction accuracy.

From a theoretic perspective, semantic transmission can be interpreted through extensions of rate-distortion theory and the information bottleneck principle \cite{tishby2000information}. This characterizes the trade-off between compression and task relevance by minimizing the mutual information between the input and its representation while preserving information that is predictive of the task output. In visual communication, this aligns naturally with the perception-distortion trade-off \cite{blau2018perception}, which demonstrates that minimizing pixel distortion alone does not guaranty perceptual or semantic fidelity. These theoretical insights motivate semantic encoding strategies that explicitly balance compression efficiency, perceptual quality, and task accuracy.

On the practical side, deep joint source-channel coding (JSCC) has enabled end-to-end semantic transmission over wireless channels \cite{bourtsoulatze2019deep}. By directly mapping visual inputs to channel symbols using neural encoders, JSCC systems bypass the conventional separation between source and channel coding, achieving graceful degradation under channel impairments. Subsequent work extends this paradigm to task-oriented settings, where semantic representations are optimized for downstream inference rather than reconstruction metrics alone \cite{ma2023task, xie2020deep}. Transformer-based  architectures further enhance the ability to capture high-level visual semantics relevant to classification, detection, and reasoning tasks \cite{wu2022vision, yoo2022real}.

Adaptive semantic transmission strategies have also been explored to dynamically allocate communication resources based on task importance and channel conditions \cite{lu2021reinforcement, wang2024adaptive}. These approaches demonstrate that semantic compression ratios and feature prioritization can be optimized in response to environmental dynamics. However, most existing task-oriented semantic communication frameworks assume predefined or static task objectives during training and deployment. The semantic encoder is typically optimized for a fixed task distribution, and the communication policy does not explicitly account for evolving user requirements or context-dependent semantic priorities.

\subsection{Generative and Large AI Models for Semantic Communication}

Beyond task-oriented semantic communication, recent research has shifted toward generative paradigms empowered by large AI models (LAMs). Instead of transmitting handcrafted features or task-specific embeddings, generative semantic communication uses learned data distributions to reconstruct semantically consistent content\cite{11361361}. Diffusion-based generative frameworks have demonstrated robustness against wireless channels by treating transmission noise as stochastic transformations in latent space \cite{grassucci2023generative, li2024end}. In such systems, the receiver reconstructs semantic outputs rather than exact physical signals, significantly reducing transmission overhead while preserving perceptual or task-level performance.

The emergence of foundation models further reshaped semantic communication. Large language models (LLMs) and VLMs provide unified representation spaces that bridge perception, reasoning, and generation \cite{jiang2026jsac_tutorial,guo2026largeai,liu2026wirelessagentic,zhang2025toward,11421643}. By exploiting pretrained world knowledge and cross-modal alignment capabilities, foundation models enable semantic encoding to operate at the level of meaning abstraction rather than signal approximation. LLM-enabled semantic communication frameworks have demonstrated that semantic compression can be guided by shared contextual knowledge between the transmitter and the receiver \cite{11050950,jiang2024wcm_lam_semantic}. Similarly, multimodal systems integrate vision-language priors into cross-modal semantic transmission, enhancing robustness and adaptability under different channel conditions \cite{jiang2025com_mag_mmsc}.

Despite these advances, current generative semantic communication frameworks remain largely task-static. Most existing work emphasizes generative quality, perceptual similarity, or downstream task accuracy, while overlooking dynamic intention modeling and interactive adaptation. Furthermore, foundation model-based communication systems are typically designed under idealized or centralized settings, with limited consideration of distributed wireless constraints, embodied agents, or edge-cloud co-design. These limitations motivate the transition from foundation model-based communication toward agent-driven semantic communication architectures, which will be elaborated in the following section \cite{jiang2024wcm_agent6g}.

\begin{table*}[ht]
\centering
\caption{Comparison of the Proposed Intention-Aware Semantic Transmission Framework with Representative Related Systems}
\label{table:intention_comparison}
\footnotesize
\setlength{\tabcolsep}{6pt}

\begin{threeparttable}

\begin{tabular}{ccllll}
\toprule
\textbf{Reference} & \textbf{Core Idea} &
\textbf{\shortstack{Explicit\\Intention}} &
\textbf{\shortstack{Proactive\\Planning}} &
\textbf{\shortstack{Robust\\ Resolution}} &
\textbf{\shortstack{Joint\\Task-Physical}} \\
\midrule

\multicolumn{6}{c}{\textit{Task-Oriented Semantic Communication}} \\
\midrule

\cite{bourtsoulatze2019deep} &
Deep JSCC semantic transmission &
$\times$ & $\times$ & $\times$ & $\checkmark$ (Passive) \\

\cite{ma2023task,xie2020deep} &
Task-specific semantic encoding &
$\times$ & $\times$ & $\times$ & $\checkmark$ (Passive) \\

\midrule
\multicolumn{6}{c}{\textit{Foundation-Model-Based Communication}} \\
\midrule

\cite{11050950,jiang2024wcm_lam_semantic} &
LLM-guided semantic compression &
$\times$ & $\times$ & $\times$ & $\checkmark$ (Implicit) \\

\midrule
\multicolumn{6}{c}{\textit{Agent and Collaborative Intelligence}} \\
\midrule

\cite{li2023camel,chen2024agentverse} &
Multi-agent semantic coordination &
$\times$ & $\checkmark$ (Non-transmission) & $\times$ & $\times$ \\

\cite{liu2024dylan,jiang2024wcm_agent6g} &
Agent-driven network orchestration &
$\times$ & $\checkmark$ (Network-level) & $\times$ & $\checkmark$ (Partial) \\

\cite{jiang2026agentcomm}&
Semantic agent communication &
$\checkmark$ (Input)  & $\checkmark$ (Server-side) & $\times$ (Text) & $\checkmark$ (Partial) \\

\midrule
\textbf{Ours} &
\textbf{Intention-aware semantic transmission for AI glasses} &
\textbf{$\checkmark$} (Auto) &
\textbf{$\checkmark$ (Transmission-level)} &
\textbf{$\checkmark$} &
\textbf{$\checkmark$ (Proactive)} \\
\bottomrule
\end{tabular}

\begin{tablenotes}
\footnotesize
\item $\checkmark$ denotes supported capability; $\times$ denotes unsupported.
``Passive'' indicates adaptation learned during training.
``Implicit'' denotes semantic compression without explicit intention modeling.
``Transmission-level'' indicates proactive semantic planning directly coupled with wireless transmission.
\end{tablenotes}

\end{threeparttable}
\end{table*}

\subsection{Edge-Cloud Collaborative Intelligence and Agent Communication}

Recent advances in LLM-based multi-agent systems have extended semantic communication beyond isolated encoder-decoder architectures toward distributed collaborative intelligence \cite{zhang2024generative,meng2026intellicise}. In these systems, multiple agents interact through structured communication channels to coordinate reasoning, task execution, and resource allocation across edge and cloud infrastructures.

Early frameworks such as CAMEL demonstrate how role-based communicative agents can engage in iterative dialogue to refine task solutions through structured semantic exchange \cite{li2023camel}. Similarly, AgentVerse provides a multi-stage collaboration pipeline in which agents with specialized roles jointly perform decision-making, execution, and evaluation, highlighting the critical role of explicit coordination and intention alignment in multi-agent environments \cite{chen2024agentverse}. These studies emphasize that communication among agents increasingly involves the exchange of intermediate reasoning states, contextual summaries, and task-level abstractions rather than raw data. Beyond static role assignment, dynamic orchestration mechanisms have been proposed to improve scalability and adaptability\cite{liu2024dylan}. Such architectures demonstrate how communication topology and coordination strategy directly influence collaborative efficiency and system robustness.

From a communication systems perspective, multi-agent collaboration has begun to be integrated into wireless and edge intelligence frameworks. LLM-enhanced multi-agent architectures  have been proposed to enable distributed planning, task coordination, and adaptive control across heterogeneous network nodes\cite{zhang2026towards,he2026agentic,jiang2026agentcomm,11355867}. These systems combine semantic reasoning with network-aware decision-making, illustrating how communication evolves from bit-level transmission toward  distributed intelligence. More broadly, AI-empowered wireless communication paradigms highlight the transition from signal optimization to semantic networking architectures.

Collectively, these works indicate a growing convergence between semantic communication and multi-agent collaborative intelligence. However, existing studies often treat semantic processing and agent coordination as loosely coupled components. A unified view that systematically integrates semantic abstraction, collaboration mechanisms, and edge-cloud communication constraints remains underexplored, particularly in large-scale distributed intelligent networks.

\subsection{Embodied and Wearable AI Systems}

The emergence of embodied AI systems marks a transition from disembodied language reasoning toward agents capable of continuous perception-action interaction in the physical world. Unlike purely digital agents, embodied systems integrate multimodal perception, high-level reasoning, and low-level motor control within dynamic environments, forming a closed perception-decision-action loop.

Recent advances in vision-language-action (VLA) architectures have significantly accelerated this transition. PaLM-E introduces a unified multimodal framework that directly grounds language instructions in physical sensorimotor representations, enabling embodied agents to reason across textual, visual, and robotic state inputs \cite{driess2023palme}. Similarly, RT-2 demonstrates that web-scale semantic knowledge can be transferred into robotic manipulation policies, bridging internet-trained language models and real-world action execution \cite{zitkovich2023rt2}. Beyond stationary robotics, embodied intelligence is progressively extending toward wearable and human-centered systems. Platforms such as VRKitchen integrate virtual reality interfaces with interactive simulation environments, enabling human-in-the-loop data collection and immersive task demonstrations \cite{gao2019vrkitchen}. In parallel, Habitat~3.0 expands embodied evaluation to human-robot cohabitation scenarios by incorporating virtual reality collaboration and social navigation tasks \cite{puig2023habitat3}. This framework emphasizes shared spaces, joint task execution, and socially aware navigation, reflecting the growing importance of wearable and assistive embodied systems in domestic environments.

Collectively, these developments indicate a convergence between embodied AI and wearable intelligence systems. Rather than operating as isolated robotic platforms, next generation embodied agents increasingly function as collaborative companions embedded within human environments. However, achieving seamless integration between large-scale semantic reasoning, real-time physical control, and socially adaptive behavior remains a significant challenge for embodied and wearable AI systems.

\begin{figure*}
	\centering
	{\includegraphics[width=0.75\linewidth]{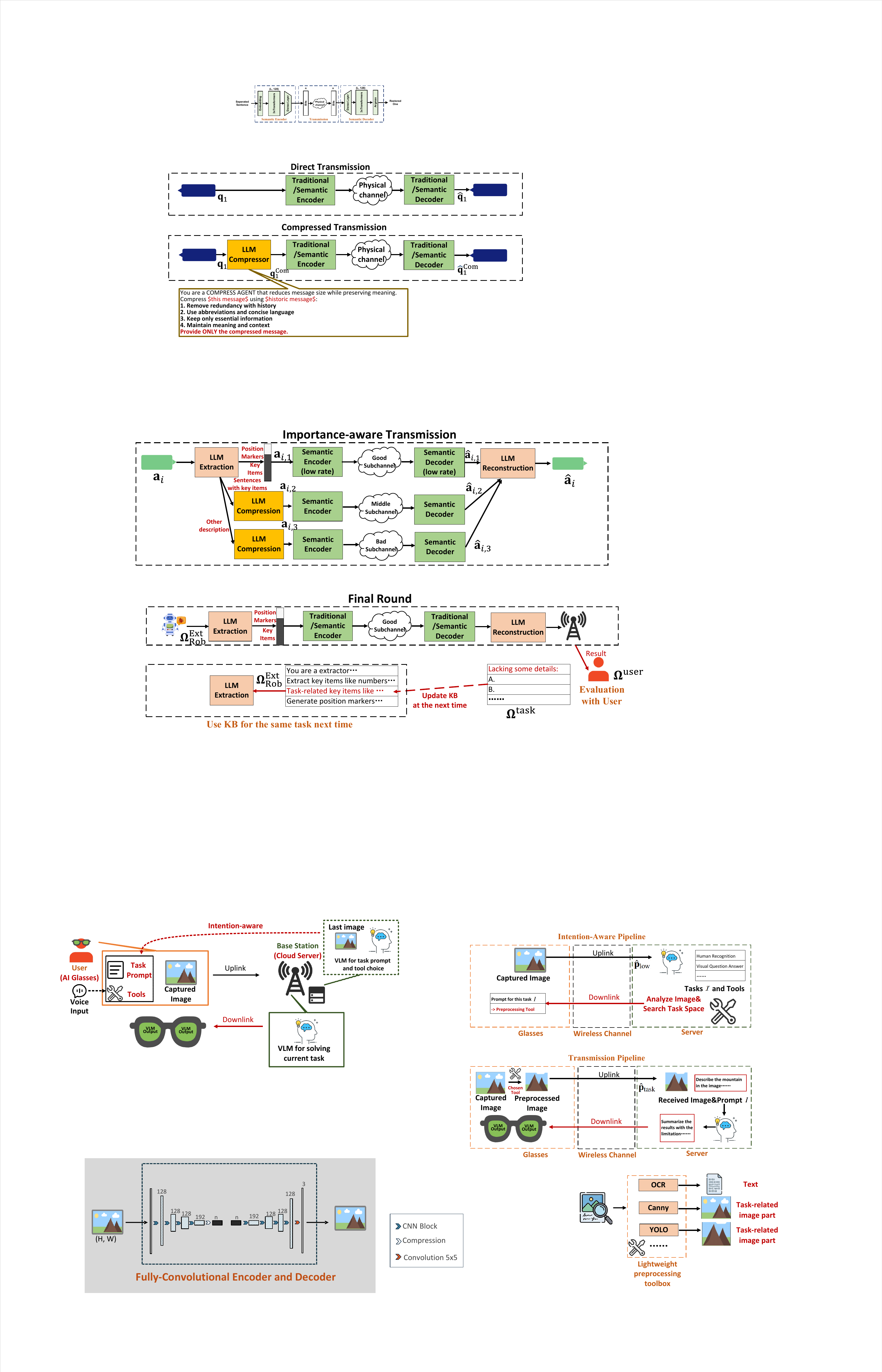}}
	\caption{Wireless communication framework between the AI glasses and the cloud server. The voice input or intention-aware control is not always activated during each transmission, because the task usually lasts for a period of time. }
	\label{Overall}
\end{figure*}
\subsection{Opportunities and Challenges}

Focusing on AI glasses, intention-aware semantic transmission enables such devices to transition from from pixel streaming toward semantic abstraction. By aligning transmitted content with inferred user intention, uplink bandwidth can be significantly reduced while preserving downstream reasoning performance. Communication costs thus scale with semantic relevance rather than spatial resolution, creating new opportunities for efficient embodied assistance under wireless and wearable constraints.

Despite this potential, several fundamental system-level challenges arise that are not addressed by existing semantic transmission frameworks.

\begin{itemize}
\item \textbf{Dynamic Intention Modeling under Wearable Constraints:}  
AI glasses infer user intention from multimodal signals such as speech commands, and contextual perception. These signals are inherently noisy and temporally evolving. Static  semantic extraction strategies cannot adapt to continuous user interaction history. Meanwhile, wearable devices operate under strict energy, computation, and latency limitations, restricting the complexity of on-device perception and control logic. A mechanism is required to dynamically model intention while maintaining lightweight edge execution.

\item \textbf{Limitation of Agent-Oriented Semantic Control and Resolution Flexibility:}  
Most existing semantic transmission pipelines are designed for fixed-resolution inputs and passive feature extraction. They do not support agent-driven prompt adaptation, tool selection, or resolution-aware semantic abstraction. In embodied wearable scenarios, visual outputs may vary across spatial scales depending on intention and task requirements. Without resolution-robust encoding and explicit separation of perception and reasoning roles across edge and cloud, semantic communication remains rigid and brittle under bandwidth fluctuations.

\end{itemize}

These challenges indicate that intention-aware semantic transmission for AI glasses cannot be achieved through conventional end-to-end compression alone. Instead, it requires: (i) explicit intention modeling as a dynamic control variable, (ii) agent-oriented edge-cloud collaborative reasoning, and (iii) resolution-robust semantic transmission mechanisms that remain robust under varying wireless and visual conditions. The structural differences between such an agent-driven framework and existing semantic transmission schemes are summarized in Table~\ref{table:intention_comparison}.

\section{System Model}

In this section, we establish the system model for intention-aware semantic agent communication for AI glasses. We first describe the wearable communication framework that connects the edge-side AI glasses to the cloud-based cognitive agent. Then, we introduce the wireless orthogonal frequency-division multiplexing (OFDM) transmission model and the effective signal-to-noise ratio (ESNR) abstraction, which serves as a bridge between physical-layer channel conditions and semantic transmission reliability.

\subsection{System Overview of Wearable Semantic AI Communication}

Wearable semantic AI systems enable real-time visual understanding through collaboration between distributed intelligent agents. As illustrated in Fig. \ref{Overall}, in the considered AI glasses scenario, the system consists of two complementary agents:

\begin{enumerate}
    \item \textbf{Edge Semantic Agent (AI Glasses):}
    The wearable device captures first-person visual observations and interacts directly with the user. Due to constraints on computational capabilities, battery capacity, and physical form factor, the glasses cannot execute large-scale VLMs locally. Instead, they perform lightweight visual preprocessing and extract task-relevant semantic information for transmission.

    \item \textbf{Cloud Semantic Agent (Cloud Server):}
    The cloud server hosts powerful VLMs and large language models (LLMs) for high-level reasoning and downstream task execution. Upon receiving the semantic information from the glasses, the cloud agent performs complex inference and generates task-level outputs.
\end{enumerate}

This edge-cloud collaboration reflects the fundamental characteristics of embodied AI systems. Unlike purely software-based agents, wearable devices must operate under strict energy budgets, real-time response requirements, and dynamic wireless environments. As a result,  reasoning is offloaded to the cloud, while the edge focuses on perception and lightweight semantic processing.

The interaction between the two agents is facilitated by wireless communication. The uplink delivers the semantic visual information extracted from the glasses, while the downlink returns concise inference results, typically in a textual format. Since the downlink payload is relatively small, this work focuses on the uplink semantic transmission problem. In particular, instead of transmitting full-resolution pixel data, the system conveys  semantic representations, whose communication efficiency and reliability are heavily dependent on wireless channel conditions.

\subsection{OFDM-Based Wireless Transmission and Effective SNR Model}

The uplink transmission from the AI glasses to the cloud agent is carried out over an OFDM-based wireless system. The OFDM system partitions a wideband signal into multiple narrowband subcarriers. Due to frequency-selective fading, different subcarriers experience heterogeneous channel gains, leading to non-uniform signal-to-noise ratios (SNRs) across subcarriers. Let $\text{SNR}_k$ denote the instantaneous SNR of the $k$-th active subcarrier, where $k = 1, \dots, K$, and $K$ is the total number of active subcarriers.

To characterize the overall channel quality under frequency-selective fading, we adopt the exponential effective SNR mapping (EESM)~\cite{sionna_phy_abstraction}, which is a widely used link-to-system level abstraction method in practical wireless systems. EESM maps the set of per-subcarrier SNRs to an equivalent effective SNR (ESNR) under an additive white Gaussian noise (AWGN) channel that yields the same block error rate (BLER) as the actual fading channel. Specifically, the ESNR is computed as
\begin{equation}
    \text{ESNR}=
    -\beta \log \left(
    \frac{1}{K}
    \sum_{k=1}^{K}
    e^{-\text{SNR}_k / \beta}
    \right),
\end{equation}
where $\beta > 0$ is a modulation- and coding-scheme-dependent parameter~\cite{sionna_phy_abstraction}. This scalar ESNR metric enables tractable link-level performance evaluation while capturing the impact of frequency selectivity.

In our semantic transmission framework, the $N_{\text{sym}}$ complex symbols generated by the semantic encoder are mapped onto the $K$ active OFDM subcarriers. In each OFDM symbol, up to $K$ semantic symbols can be transmitted in parallel across subcarriers. When $N_{\text{sym}} > K$, multiple OFDM symbols are employed, and the semantic symbols are sequentially scheduled over consecutive OFDM symbols. This mapping explicitly connects the semantic encoder output to the physical-layer OFDM structure.

In the considered wearable semantic communication system, the ESNR serves as a unified physical-layer reliability indicator that interfaces with the semantic transmission process. For a given modulation and coding scheme, the ESNR determines the BLER of each OFDM frame carrying semantic symbols. Transmission errors at the physical layer may corrupt the transmitted semantic representations, thereby degrading the semantic reconstruction quality at the cloud agent. Consequently, the ESNR indirectly affects the accuracy of downstream AI tasks, establishing a cross-layer link between wireless channel conditions and semantic-level performance.

\begin{figure*}
	\centering
\subfigure[]{\includegraphics[width=0.9\linewidth]{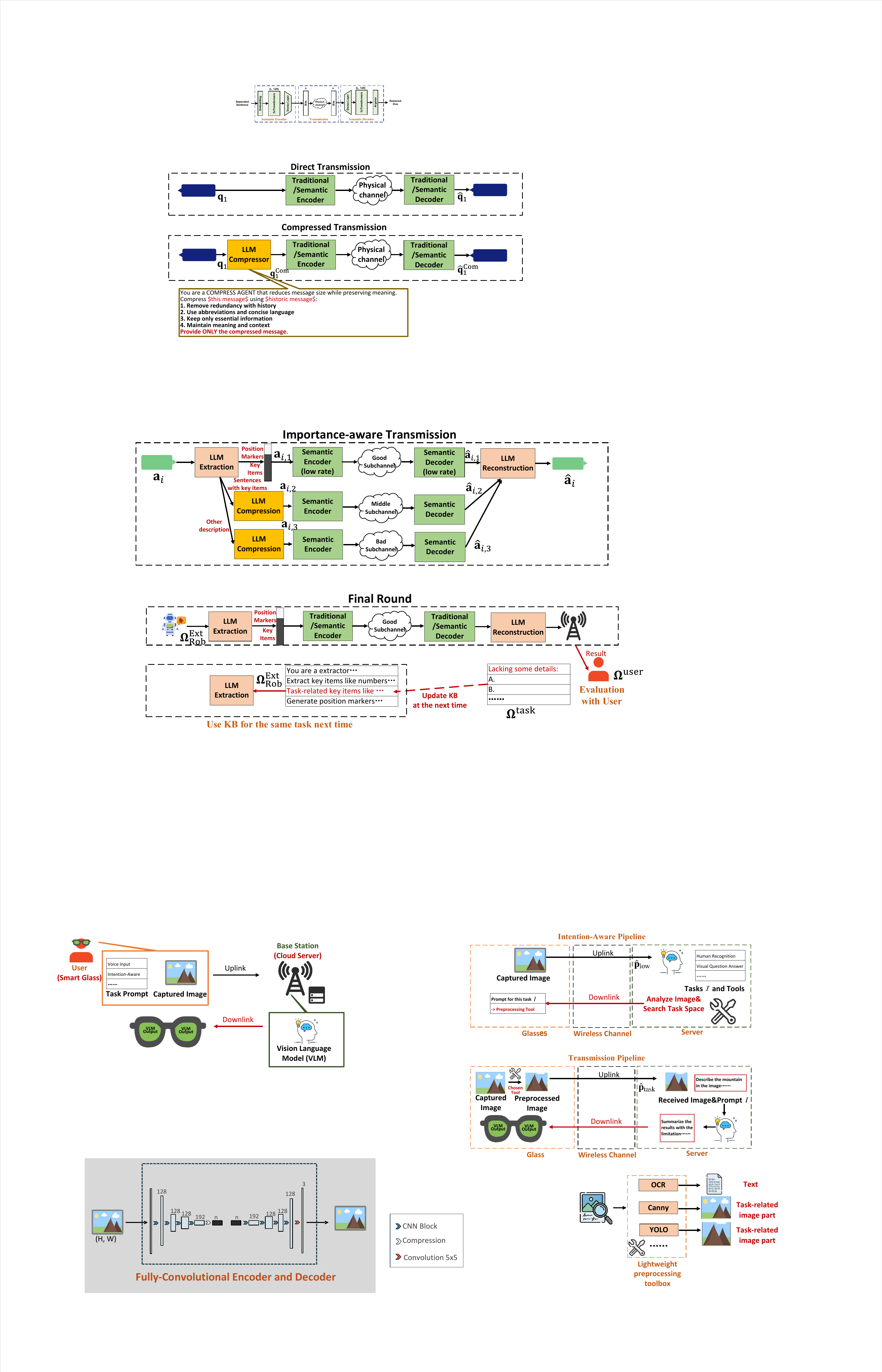}}
\subfigure[]{\includegraphics[width=0.9\linewidth]{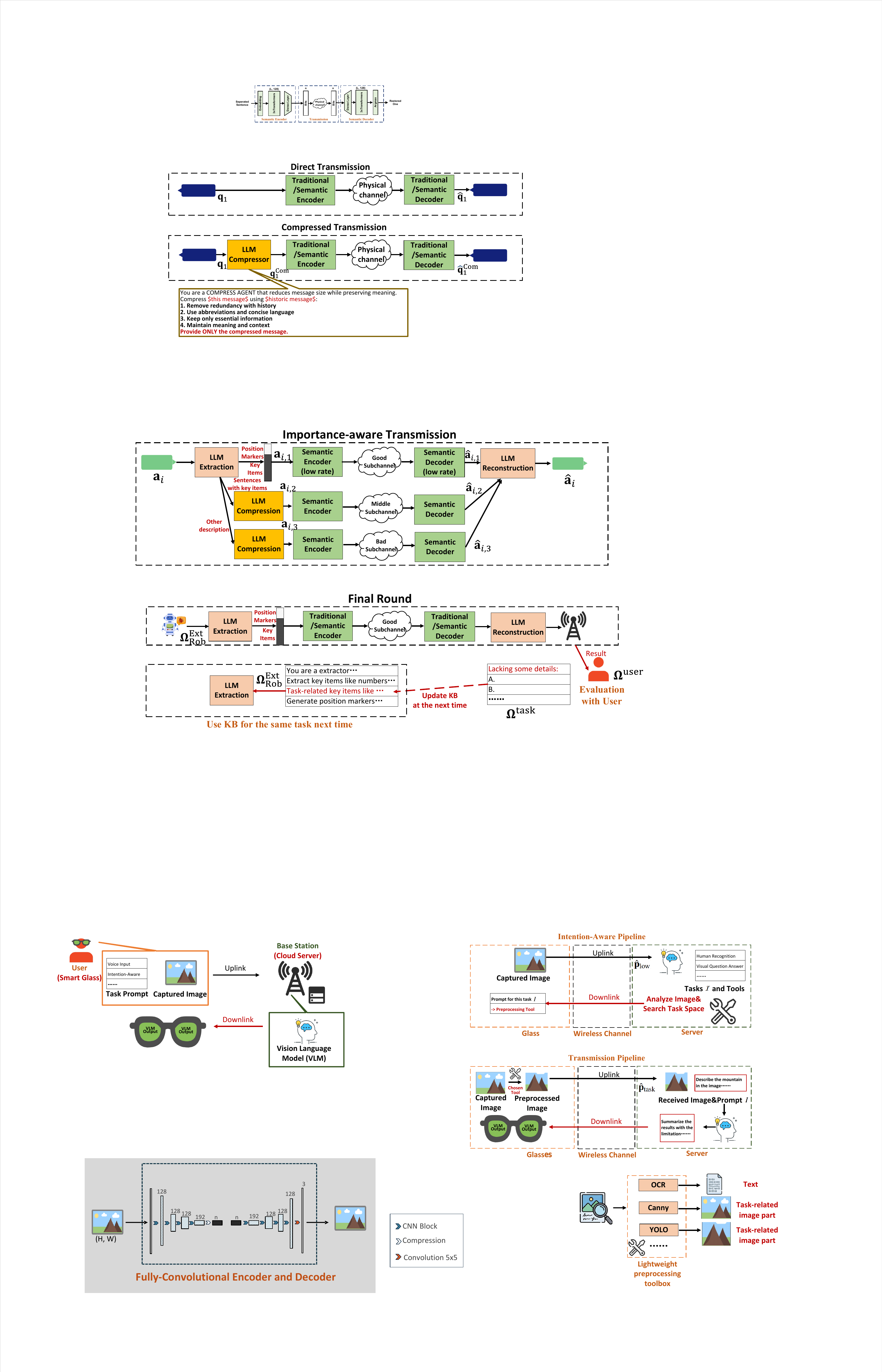}}
	\caption{(a) The proposed intention-aware pipeline with the task recognition at the cloud server. (b) Using the chosen preprocessing tools for the subsequent transmissions.} 
	\label{intent_trans}
\end{figure*}

\section{Proposed Framework for Semantic Agent Communication for AI Glasses}
\label{s3}

This section presents the communication framework between the server and the AI glasses. To improve transmission efficiency and reduce user intervention, user intention is first inferred. Then, the communication overhead is further reduced through a semantic preprocessing tool deployed on the glasses. Meanwhile, a robust semantic encoder-decoder is also applied to the processed image.

\subsection{System Architecture and Design Principle}

We consider a semantic communication framework consisting of an edge agent deployed on AI glasses and a cloud server connected through a wireless channel. 
The system is designed to transmit task-relevant semantic information rather than full-resolution raw data. The uplink transmission from the glasses to the server consists of three modules: the intention-aware pipeline, semantic preprocessing, and the semantic encoder-decoder.

The intention-aware transmission operates in two stages: 
(i) a lightweight low-resolution transmission for intention recognition, and 
(ii) a task-specific semantic transmission for high-fidelity reconstruction. 
In the following, we first describe the second stage, i.e., the task-specific semantic transmission process.

Let $\mathbf{p} \in \mathbb{R}^{H \times W \times 3}$ denote the captured RGB image. 
Unlike continuous video streaming systems, practical AI glasses are typically subject to stringent energy constraints. High-frequency frame acquisition and transmission would incur substantial sensing, encoding, and wireless transmission overhead. Hence, many wearable vision systems adopt an event-driven or low-duty-cycle image capture mechanism. Following this design principle, we model the uplink transmission based on intermittently captured static images.  The received image can be expressed as
\begin{equation} 
\widehat{\mathbf{p}}_{\rm task}= {\tt SC}_{\rm de}(\mathcal{C}({\tt SC}_{\rm en}(g_{\rm pre}(\mathbf{p})))), \label{eq2}
\end{equation}
where $g_{\rm pre}(\cdot)$ denotes the lightweight preprocessing module on the glasses,
$\mathcal{C}(\cdot)$ models the wireless channel,
and  ${\tt SC}_{\rm en}(\cdot)$ and ${\tt SC}_{\rm de}(\cdot)$ denote the semantic encoder on the glasses and the semantic decoder at the server, respectively.

Then, the received image $\widehat{\mathbf{p}}_{\rm task}$ is used for downstream tasks $\mathcal{T}(\cdot)$ through the VLM agent at the server. 
The system performance is evaluated at the task level via the loss
\begin{equation}
\ell\big(\mathcal{T}(\mathbf{p}),\mathcal{T}(\widehat{\mathbf{p}}_{\rm task})\big), \label{eq3}
\end{equation}
where $\ell(\cdot,\cdot)$ denotes a task-dependent loss function, which is chosen according to the specific scenario.

Compared to the uplink, the downlink from the server to the glasses only carries the result from the server in text format, and the downlink resources are usually efficient. Thus, downlink transmission overhead is not the focus of this study.

\subsection{Intention-Aware Pipeline}

The AI glasses support two task activation mechanisms: explicit voice triggering and implicit intention-aware recognition. 
In the explicit mode, users directly specify the desired task via voice commands, upon which the corresponding task-specific transmission strategy is activated.

In addition to explicit interaction, the system operates in an intention-aware mode, illustrated in Fig.~\ref{intent_trans}(a). In this mode, the glasses periodically captures a low-resolution image (256 $\times$ 256) for intention probing. 
Let $\mathbf{p}_{\rm low}$ denote the downsampled image. 
After uplink transmission, the server receives $\widehat{\mathbf{p}}_{\rm low}$ and predicts the task label using a VLM,
\begin{equation}
 I = f_{\rm VLM}(\widehat{\mathbf{p}}_{\rm low}),
\end{equation}
where $I$ is the current intention predicted by the VLM. The VLM can search the task space $\mathcal{I} = \{\text{text-reading}, \text{document}, \text{scene}, \ldots\}$, which denotes the stored task commands. Also, if the predicted intention cannot be found $I \notin \mathcal{I}$, the new intention will be added into the task space with the guidance of the user.
This low-resolution probing stage enables intention inference with limited bandwidth consumption.

Once a stable task label $I$ is obtained, the system switches to the task-specific transmission pipeline shown in Fig.~\ref{intent_trans}(b), where preprocessing and semantic encoding strategies are selected according to $I$. 
During task execution, the consistency between the reconstructed image $\widehat{\mathbf{p}}_{\rm task}$ and the current task $I$ is continuously evaluated. 
Specifically,
\begin{equation}
    \delta = f'_{\rm VLM}(\widehat{\mathbf{p}}_{\rm task}, I),
\end{equation}
where $\delta \in \{0,1\}$ is a binary consistency indicator. 
Here, $\delta = 1$ indicates that the received content remains compatible with the intended task, while $\delta = 0$ indicates a potential intention change.
The function $f'_{\rm VLM}(\cdot)$ shares the same backbone model as $f_{\rm VLM}(\cdot)$ but performs conditional verification given the task $I$, rather than label prediction through different prompts.
If $\delta = 0$, the system terminates the task-specific transmission and reverts to the low-resolution intention recognition stage in Fig.~\ref{intent_trans}(a). 
This mechanism forms a closed-loop control structure that enables adaptive switching between intention detection and task-oriented transmission. Most importantly, the intention evaluation can be activated once per second because the task cannot be switched frequently. The cost of generating the indicator  $\delta $ can be omitted in this case.

Overall, the proposed intention-aware pipeline provides a unified framework that integrates explicit voice-triggered tasks and implicit visual intention inference. 
By coupling adaptive transmission control with continuous intention monitoring, the system achieves efficient bandwidth utilization while remaining responsive to spontaneous changes in user intention.
\begin{figure}[!h]
	\centering
	{\includegraphics[width=0.99\linewidth]{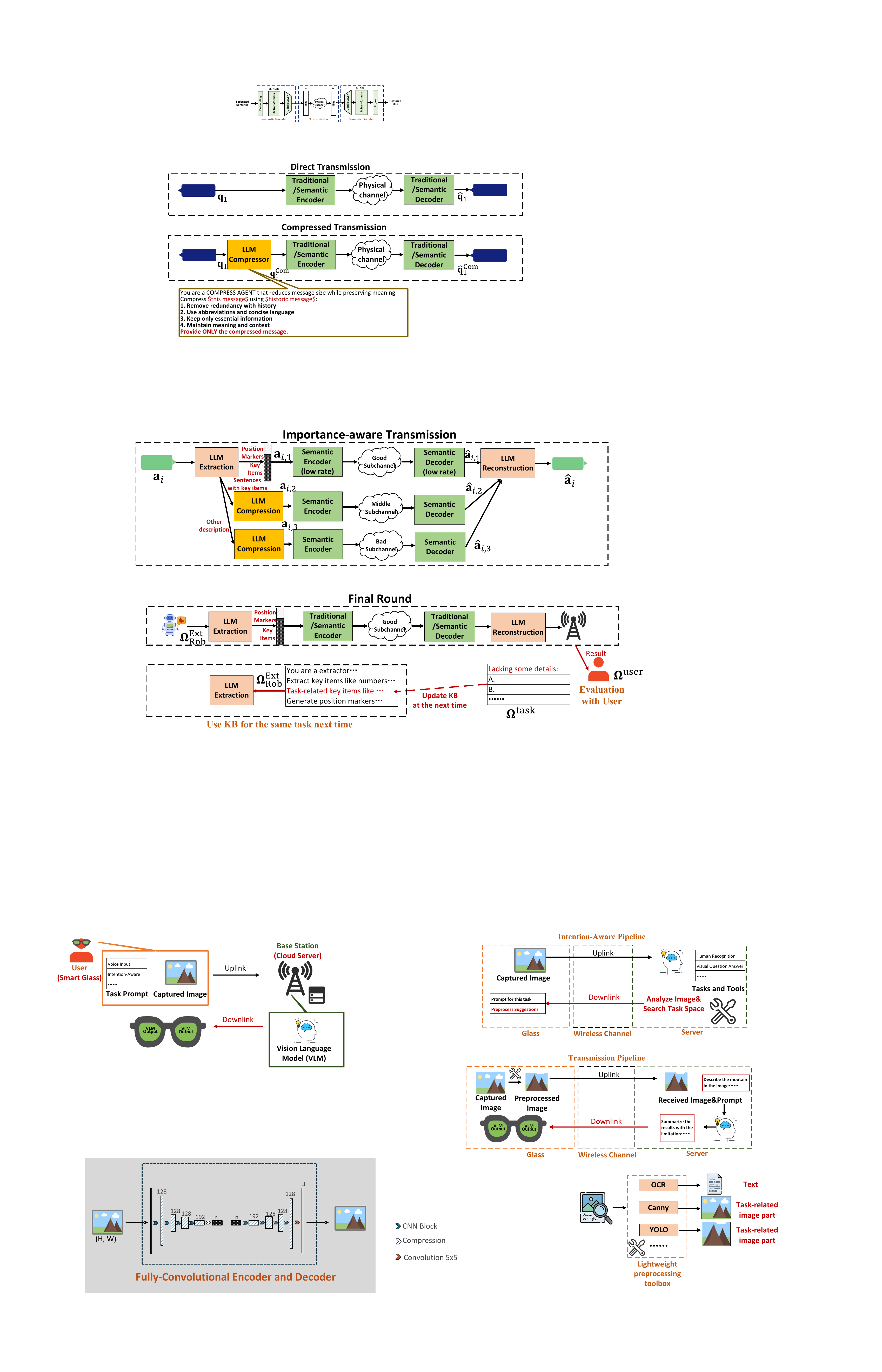}}
	\caption{Three preprocessing tools  for different scenarios $I$. More tools are also acceptable and VLM will select the appropriate one based on the intention.}
	\label{tool}
\end{figure}

\begin{figure*}
	\centering

\includegraphics[width=0.9\linewidth]{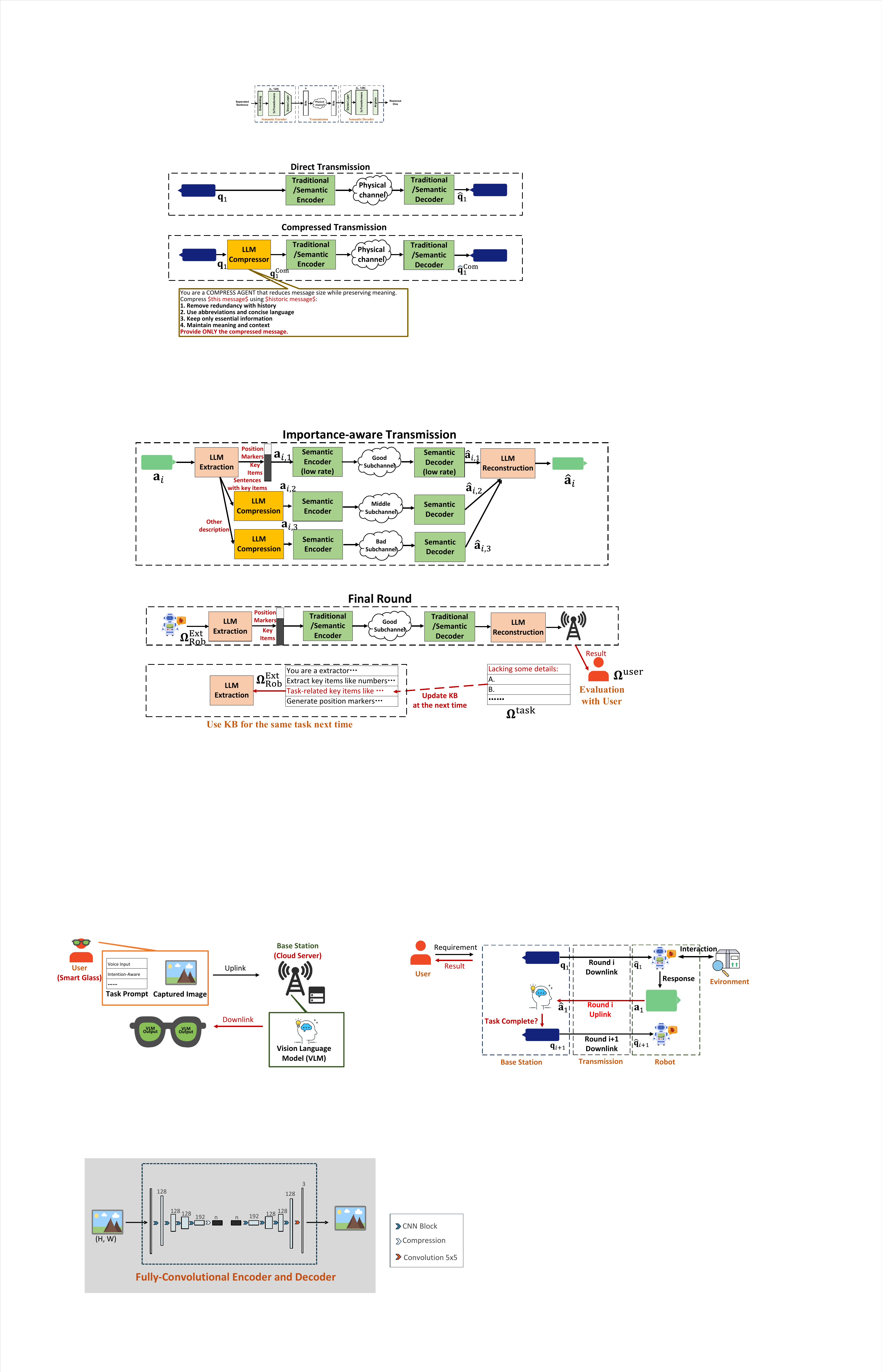}
	\caption{Robust semantic encoder-decoder in a fully convolutional architecture, where the hidden channel dimensions are 128 and 192. } 
	\label{Transarchi}
\end{figure*}
\subsection{Lightweight Semantic Preprocessing Tools }

To reduce uplink transmission overhead while preserving task-relevant information, we deploy multiple lightweight semantic preprocessing modules on the glasses.
Specifically, the preprocessing tool is selected from a candidate set. Fig. \ref{tool} illustrates several lightweight tools that can be easily applied on the glasses, yielding
\begin{equation}
\mathcal{G}_{\rm pre} = \left\{ g_{\rm pre}^{\rm OCR}, \; g_{\rm pre}^{\rm Canny}, \; g_{\rm pre}^{\rm YOLO}, \ldots \right\}
\end{equation}
where the tools are prepared for  different scenarios.
Given the recognized intention $I \in \mathcal{I}$, the preprocessing function is adaptively selected according to the VLM's analysis.

The three candidate tools and their corresponding scenarios are described as follows:

\textbf{1) Text-centric scenarios: $g_{\rm pre}^{\rm OCR}(\cdot)$.}
For tasks where textual semantics dominate the scene, such as real-time translation, document reading, note summarization, and writing assistance, transmitting full-resolution images is inefficient.
In these scenarios, we apply optical character recognition locally on the glass:
\begin{equation} 
\mathbf{p}_{\rm task}= g_{\rm pre}^{\rm OCR}(\mathbf{p}) = \mathbf{p}_{\rm text},
\end{equation}
where $\mathbf{p}_{\rm text}$ denotes the extracted textual tokens or a structured text representation. Since OCR-based preprocessing produces compact textual tokens $\mathbf{p}_{\rm text}$ with negligible bandwidth consumption, conventional digital transmission is sufficient and its encoding procedure is omitted.
We therefore focus on the image-domain semantic representation.

For example, when a user looks at a restaurant menu and requests translation, the OCR module extracts the menu text, and only the recognized text is forwarded to ${\tt SC}_{\rm en}(\cdot)$.
Compared to transmitting $\mathbf{p} \in \mathbb{R}^{H \times W \times 3}$, sending $\mathbf{s}_{\rm text}$ significantly reduces pixel-level semantic redundancy.

\textbf{2) Document scenarios: $g_{\rm pre}^{\rm Canny}(\cdot)$.}
For visually grounded reasoning tasks where structural layout is more critical, we adopt a Canny edge detector:
\begin{equation}
\mathbf{p}_{\rm task}=g_{\rm pre}^{\rm Canny}(\mathbf{p}) = \mathbf{p}_{\rm doc},
\end{equation}
where $\mathbf{p}_{\rm doc}$ denotes the document region of interest (ROI) localized via Canny-based boundary detection. 

For instance, when a user asks questions about a plotted curve or a flowchart, the structural contours are more important than background details. The edge map suppresses redundant visual content while preserving geometric relationships, leading to improved semantic efficiency before transmission.

\textbf{3) Object scenarios: $g_{\rm pre}^{\rm YOLO}(\cdot)$.}
For tasks focusing on object recognition, scene description, or product identification, object semantics are most relevant.
We therefore deploy a lightweight YOLO detector on the glass:
\begin{equation}
\mathbf{p}_{\rm task}=g_{\rm pre}^{\rm YOLO}(\mathbf{p}) = \mathbf{p}_{\rm obj},
\end{equation}
where $\mathbf{p}_{\rm obj}$ denotes an image crop that contains all detected objects.

As an example, when the user looks at a shelf and requests product identification, the YOLO module extracts object proposals locally.
Instead of transmitting the entire frame, the glasses can send the cropped image part, which is further processed by the server.

Overall, the tool selection of $g_{\rm pre}^{\alpha}(\cdot)$ transforms the raw image $\mathbf{p}$ into a compact semantic representation before semantic encoding and wireless transmission in \ref{eq2}. This task-adaptive semantic reduction effectively lowers the entropy of transmitted content while maintaining performance in the task-level objective defined in \ref{eq3}. 
Combined with the intention-aware initiation mechanism, the proposed preprocessing module enables a hierarchical edge-cloud collaboration framework that balances local computation and uplink bandwidth efficiency. However, compared to the raw captured image from the glasses, the processed one has different resolutions and the design of the semantic coding should be further investigated.

\subsection{Resolution-Robust Semantic Encoder-Decoder}

After lightweight preprocessing, the uplink image transmission is performed by the semantic encoder-decoder pair ${\tt SC}_{\rm en}(\cdot)$ and ${\tt SC}_{\rm de}(\cdot)$.
The semantic encoder and decoder are instantiated using the fully convolutional transforms proposed in~\cite{begaint2020compressai} and initialized with pretrained weights. To adapt to varying input resolutions, the robust semantic encoder-decoder is implemented through the following steps.

Given $\mathbf{p}_{\rm task}$ with spatial size $H \times W$, reflection padding is first applied to the nearest multiple of $64$:
\begin{equation} 
\widetilde{H}=\lceil H/64 \rceil \times 64, 
\quad 
\widetilde{W}=\lceil W/64 \rceil \times 64,
\end{equation}
where $\widetilde{\mathbf{p}} \in \mathbb{R}^{3 \times \widetilde{H} \times \widetilde{W}}$.
The encoder ${\tt SC}_{\rm en}(\cdot)$ consists of four stages of $5\times5$ convolutions with stride~2 followed by GDN activation. 
The first three stages maintain 128 channels, and the final stage projects to 192 channels without activation, resulting in an overall spatial downsampling factor of $16$. 
The latent representation is
\begin{equation}
\mathbf{y} = g_a(\widetilde{\mathbf{p}}) 
\in \mathbb{R}^{M \times \frac{\widetilde{H}}{16} \times \frac{\widetilde{W}}{16}},
\end{equation}
with a total number of elements
\begin{equation}
L = M \cdot \frac{\widetilde{H}}{16} \cdot \frac{\widetilde{W}}{16}.
\end{equation}
The decoder ${\tt SC}_{\rm de}(\cdot)$ mirrors this structure using four transposed convolutions with $5\times5$ kernels and stride~2 together with IGDN activation, following the channel mapping $192 \rightarrow 128 \rightarrow 128 \rightarrow 128 \rightarrow 3$. 
The reconstructed output is cropped back to $H \times W$ and clamped to $[0,1]$.

To enable bandwidth-adaptive transmission, we introduce a compression ratio $n \in \{2,4,8,16\}$ applied to the latent tensor before transmission.
Each latent element is uniformly quantized to
\begin{equation}
b = \frac{32}{n} ~~\text{bits}.
\end{equation}
Let $y_{\min}$ and $y_{\max}$ denote the global minimum and maximum of $\mathbf{y}$.
The quantized latent $\mathbf{y}_q$ is obtained as
\begin{equation}
\mathbf{y}_q =
\left\lfloor
\frac{\mathbf{y} - y_{\min}}{y_{\max} - y_{\min}} \cdot (2^b - 1)
\right\rceil
\cdot \frac{y_{\max} - y_{\min}}{2^b - 1}
+ y_{\min},
\end{equation}
where $\lfloor \cdot \rceil$ denotes rounding to the nearest integer.

At the transmitter, $n$ quantized elements (each represented with $b$ bits) are packed into one $32$-bit real symbol, and two real symbol uses are combined into one complex symbol $s$.
Denoting the transmitted symbol sequence by $\mathbf{s}$, the total number of complex channel uses is
\begin{equation}
N_{\rm sym} = \left\lfloor \frac{L}{2n} \right\rfloor.
\end{equation}

Under the OFDM-based physical layer model described in Section~II, 
each OFDM symbol occupies $K$ orthogonal subcarriers and thus carries $K$ parallel complex symbols. 
Therefore, transmitting $N_{\rm sym}$ complex symbols requires
\begin{equation}
N_{\rm OFDM} = \left\lceil \frac{N_{\rm sym}}{K} \right\rceil
\end{equation}
OFDM symbols. 
If $N_{\rm sym}$ is not an integer multiple of $K$, zero-padding is applied to the last OFDM symbol.

Increasing $n$ reduces the number of transmitted symbols proportionally while decreasing quantization resolution.
Specifically, $n=2$ allocates $16$ bits per element, $n=4$ allocates $8$ bits, and $n=8$ allocates $4$ bits.
As an illustrative example, for a cropped image size $704 \times 1024$ (already multiples of $64$ in each dimension), the encoder produces a latent tensor of size $192 \times 44 \times 64$, giving $L = 540{,}672$. 
With $n=4$, the number of transmitted complex symbols becomes 
\( N_{\rm sym} = \lfloor 540{,}672 / 8 \rfloor = 67{,}584 \), 
achieving a $4\times$ bandwidth reduction compared to transmitting the latent in raw 32-bit floating-point format.

Overall, the fully convolutional architecture ensures robustness to varying spatial resolutions and heterogeneous semantic inputs.  The  quantization $b$ provides explicit control between reconstruction performance and channel usage through the transmitted symbol stream.

\section{Simulation Results}

In this section, we evaluate the performance of the proposed intention-aware semantic communication framework for AI glasses communication systems. We compare the proposed intention-driven strategy with conventional full-resolution image transmission schemes in terms of bandwidth consumption and task accuracy. Furthermore, the simulation results reveal the fundamental trade-off among semantic resolution, transmission reliability, and task accuracy, thereby validating the effectiveness of the proposed  semantic agent communication design.

\subsection{Cases, Datasets, and Evaluation Metrics}
\label{subsec:cases_datasets_metrics}

We consider three representative scenarios of increasing 
semantic and transmission complexity. In all cases, the AI glasses perform lightweight edge processing, the processed result is transmitted over a 
noisy wireless channel, and a server-side VLM (GPT-4o) 
performs downstream reasoning. 

For typical mobile processors, OCR and Canny-based edge extraction incur on-device delays are commonly around 10 ms, while lightweight YOLO inference typically requires 30-60 ms depending on edge device, such as Nvidia Jetson \cite{shin2022deep}. 
Transmission latency is proportional to the semantic payload size and channel coding rate, and can be significantly reduced under low-resolution semantic abstraction. Cloud-side VLM reasoning dominates the latency budget and usually ranges from hundreds of milliseconds to around one second, depending on API configuration\cite{xu2026poster}.  In general, the transmission latency is not the key issue because the user still waits for the response of the VLM. In contrast, the transmission overhead and the edge computation capability limit the mobile processors.

\textbf{Case I: Text Reading and Answering.}

\begin{itemize}
    \item \textbf{Scenario Description:}
In this scenario, no image data are transmitted. The  glass applies 
an on-device OCR engine \cite{opencv} to extract legible text from the captured frame 
and transmits only the resulting UTF-8 string. The server performs 
question answering solely based on the recovered text. 
This configuration minimizes bandwidth consumption but discards 
all non-textual visual information.

\item \textbf{Dataset Description:}
Experiments are conducted on 300 receipt samples and 300 document images from the DocVQA benchmark~\cite{mathew2021docvqa}. To emulate wearable first-person capture, each image is augmented with 30\% background padding, additive sensor noise, and random lighting perturbation.

\item \textbf{Evaluation Metrics:} 
Task performance is measured by answer accuracy, defined as the 
fraction of samples for which the predicted answer exactly matches 
the ground-truth answer, i.e., task success rate (SR).
\end{itemize}

\textbf{Case II: Document Reading and Answering.}

\begin{itemize}
\item\textbf{Scenario Description:}
This scenario targets document-centric transmission. The AI glasses apply Canny edge detection~\cite{opencv} to localize the dominant 
rectangular document boundary and crops the frame to this region 
before transmission. Only the cropped document ROI is transmitted 
to the server, which performs document-level question answering.

\item \textbf{Dataset Description:}
The same 300 DocVQA images used in Case~I are adopted. 
Photo-capture augmentation (background padding, 
sensor noise, and lighting perturbation) is applied 
prior to Canny-based boundary detection and preprocessing.

\item \textbf{Evaluation Metrics:}
Performance is also evaluated using SR, defined as 
the proportion of correctly answered questions in this case.
\end{itemize}

\textbf{Case III: Scene Watching.}
\begin{itemize}
    \item \textbf{Scenario Description:}
This scenario considers general scene understanding. The AI glasses run YOLOv8n~\cite{yolov8} with a confidence threshold of 0.30 to detect 
objects in the captured frame. All detected bounding boxes are merged 
into a single minimal enclosing rectangle with an additional 5\% margin. 
Only this unified crop is transmitted to the server for scene-level reasoning.

\item \textbf{Dataset Description:}
Experiments use 300 images from the COCO dataset~\cite{lin2014microsoft}, 
restricted to samples containing at least four objects across at least 
two distinct categories to ensure sufficient semantic diversity. 
No additional photo-capture augmentation is applied.

\item 
\textbf{Evaluation Metrics:}
Two complementary metrics are reported:

\emph{Object Coverage}\cite{yang2025thinking}: the fraction of ground-truth COCO categories 
mentioned in the VLM output, with synonym expansion bridging dataset 
labels and natural-language expressions.

 \emph{BERTScore}~\cite{bertscore1,bertscore}: the maximum token-level cosine 
similarity between the VLM response and the five human-written COCO captions.

\end{itemize}

Each case employs task-specific evaluation metrics aligned with its semantic abstraction level. Case I and Case II focus on discrete question answering or object identification, where task success rate (SR) provides a direct measure of semantic correctness. Case III involves open-ended scene understanding, where semantic coverage and language similarity metrics (e.g., Object Coverage and BERTScore) are more appropriate to quantify descriptive fidelity. Although the metrics differ across cases, they consistently reflect semantic efficiency, defined as task-relevant information preservation under constrained transmission resources.

We consider an uplink OFDM transmission over a frequency-selective 
Rayleigh fading channel. Unless otherwise specified, the number of 
subcarriers is set to $K=64$, and 16-QAM modulation is adopted. 
Channel coding is implemented using an LDPC code with rate $1/2$. Least squared channel estimation is assumed at the receiver.
\textbf{For image-based cases (Case~II and Case~III)}, we compare:
\begin{itemize}
\item JPEG+LDPC  transmission on the full frame,
\item JPEG+LDPC  transmission on the preprocessed image,
\item Semantic  transmission on the preprocessed image with 
 $b$ bits for each latent element.
\end{itemize}

\subsection{Performance of Case I: Text Reading and Answering}

\begin{table}[h]
    \centering
    \begin{threeparttable}
    \caption{SR and bandwidth for OCR text and JPEG image transmission.}
    \footnotesize
    \label{Metric1}
    \begin{tabular}{lccc}
        \toprule
        Dataset  & OCR Text & JPEG& JPEG (quality=50)\\
        \midrule
        Receipt (SR)  & 90\% & \textbf{100\%} & 98\%\\
        Receipt, HighRes (SR)  & 100\% & \textbf{100\%}&100\% \\
        Doc (SR)  & 44\% & \textbf{78\%} & 73\% \\
        \midrule
         Average Bandwidth (KBytes) & \textbf{0.13} & 81 & 24\\
        \bottomrule
    \end{tabular}
    \begin{tablenotes}
        \footnotesize
        \item High resolution (HighRes) receipts are obtained by filtering out blurry images.\cite{pertuz2013analysis}
    \end{tablenotes}
    \end{threeparttable}
    
\end{table}

In Case I, the OCR-based transmission and full-image transmission are compared, in which the quality of the original image and the OCR result affect the task performance.
Table~\ref{Metric1} compares the SR and average transmission 
bandwidth for text transmission only with OCR and full JPEG image transmission. 
A substantial reduction in bandwidth is observed when transmitting only OCR text, with the average payload decreasing from 81~KBytes to 0.13~KBytes. The JPEG with the optimized quality (50) can also reduce the bandwidth into 24~KBytes but lose the accuracy under some blurry images.
For receipt images, OCR-based transmission achieves a 90\% SR, only 10 percentage points lower than full image transmission, indicating that the task is predominantly text-centric and can be effectively supported without visual content beyond recognized characters. In contrast, for general document images, the SR drops from 78\% (JPEG image) to 44\% (OCR text). This performance degradation suggests that 
document reasoning often relies on layout structure, tables, or non-textual visual cues that are lost in pure text extraction. These results highlight that the  OCR-only transmission offers extreme bandwidth efficiency but its achievable task performance depends strongly on whether the downstream reasoning task is purely textual or requires structured visual information.
\begin{figure}[h]
	\centering

	{\includegraphics[width=0.9\linewidth]{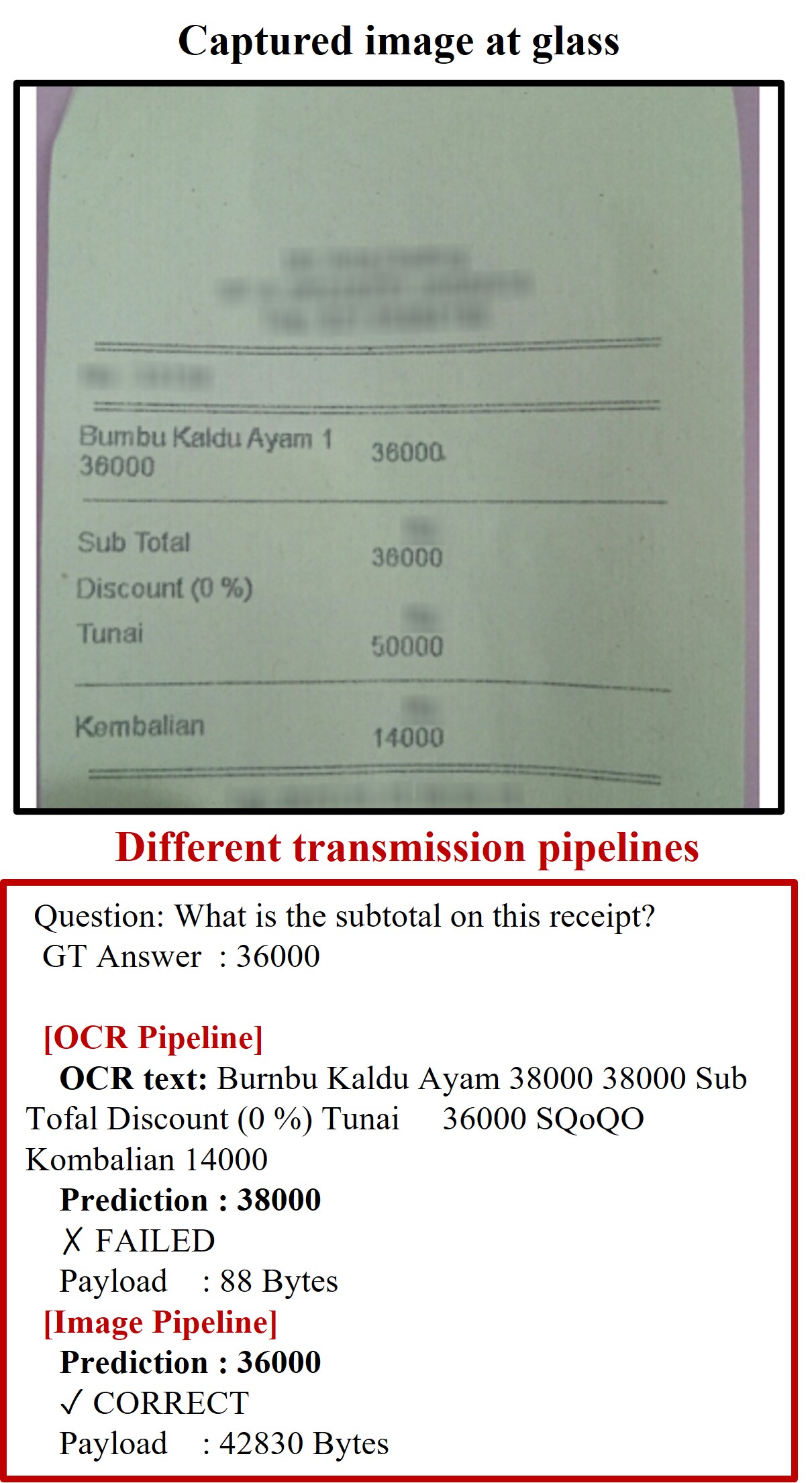}}

	\caption{Failed example in Case I, where the low-resolution image reduces OCR accuracy.}
	\label{EX_1}
 
\end{figure}

The different SRs of the receipt dataset can be easily recovered if we only use the OCR preprocessing under high-resolution images. Here, blurry images can be easily removed through quick measurement, such as the variance of the Laplacian \cite{pertuz2013analysis}. The images with the 10\% worst sharpness scores were excluded from the OCR preprocessing. Then, the SR goes back to 100\% without too much bandwidth increase. This can be regarded as a simple trick to fully exploit the advantages of the preprocessing tools and  the detailed differences are illustrated in the following example.

In the failure example shown in Fig. \ref{EX_1}, 
although this value appears in the 88-byte OCR text payload, the VLM predicts 38000, resulting in task failure. The error arises from incorrect character recognition. Multiple numerical values (36000 and 38000) are presented in the OCR output and without two-dimensional structural cues linking numbers to their corresponding semantic fields, the model cannot repair the incorrect values. In contrast, when the full JPEG image (42.8~KBytes) is transmitted, the VLM can exploit visual layout, text positioning, and contextual formatting to correctly associate the subtotal label with the corresponding value. This example indicates that layout-preserving visual information can be critical for reasoning, despite the substantially higher bandwidth cost.

Overall, Case~I demonstrates that OCR-only transmission can achieve extreme bandwidth efficiency, reducing payload size by several orders of magnitude compared to image transmission. For purely text-centric tasks such as simple receipts, this approach remains largely effective. However, its performance degrades when the task requires layout-aware or structurally grounded reasoning. These results suggest that textual completeness alone does not guarantee semantic sufficiency, which motivates the exploration of 
layout-preserving visual transmission in subsequent cases.

\begin{figure*}[!h]
	\centering

		\subfigure[]{\includegraphics[width=0.45\linewidth]{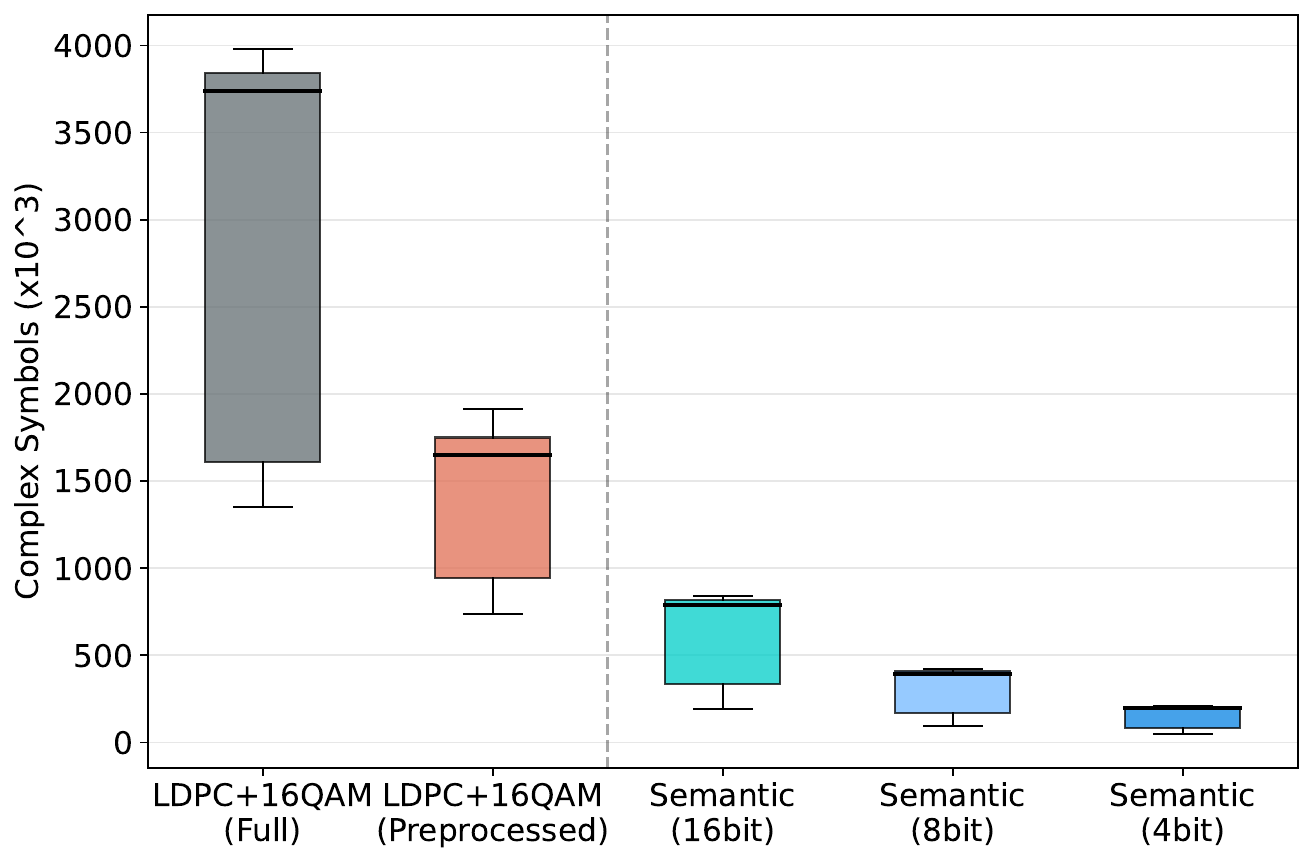}}
        \subfigure[]{\includegraphics[width=0.45\linewidth]{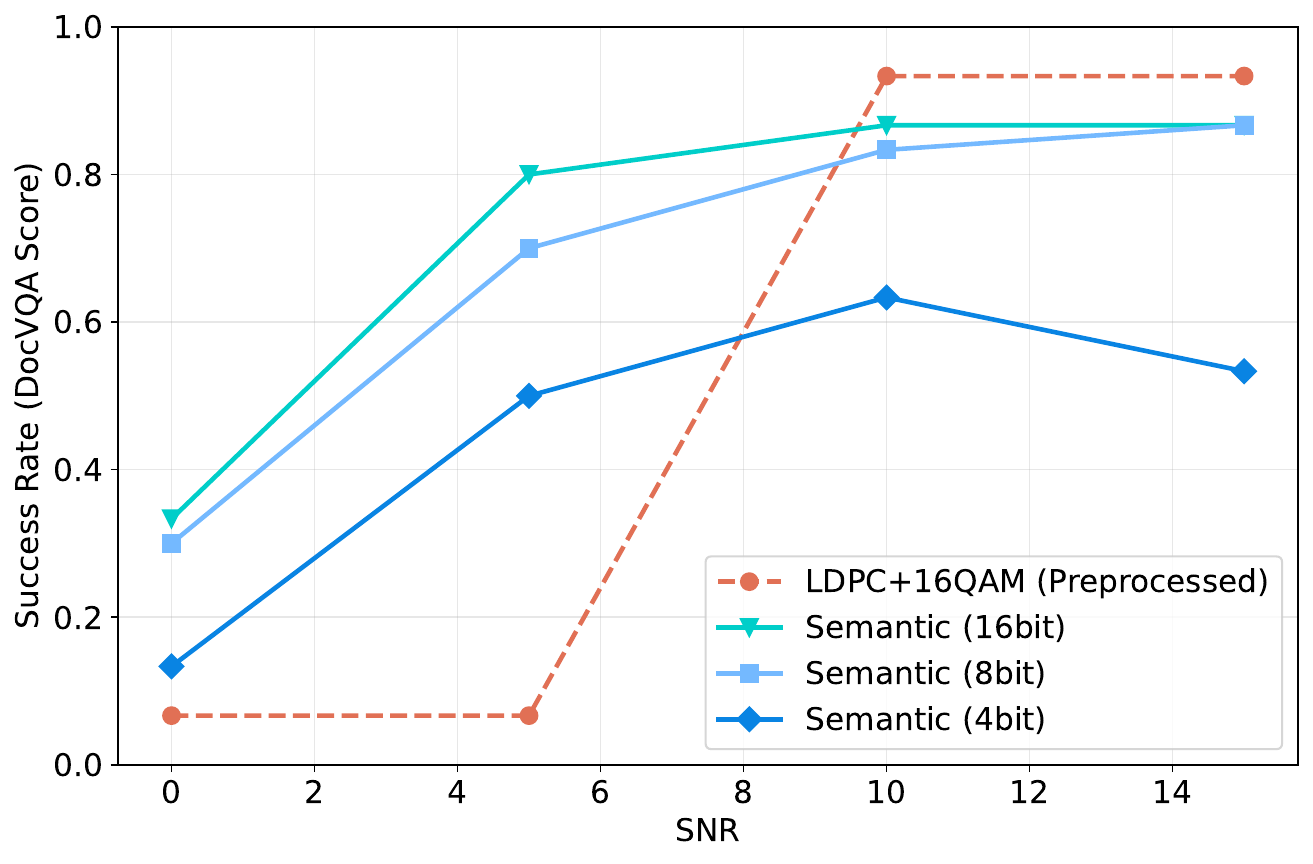}}
	\caption{(a) Bandwidth comparison of the different document transmission methods. (b) SR under different preprocessing and transmission schemes for document images under varying channel conditions. }
	\label{case2}
\end{figure*}
\begin{figure*}[!h]
	\centering

		\subfigure[]{\includegraphics[width=0.23\linewidth]{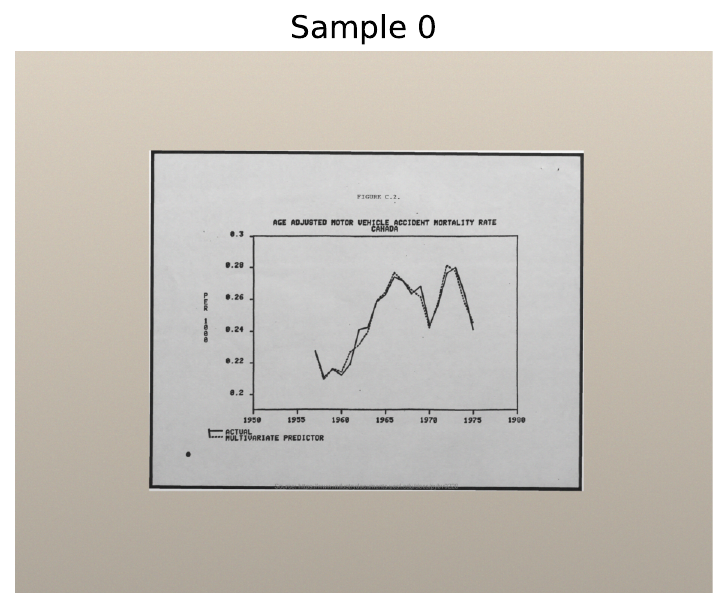}}
        \subfigure[]{\includegraphics[width=0.245\linewidth]{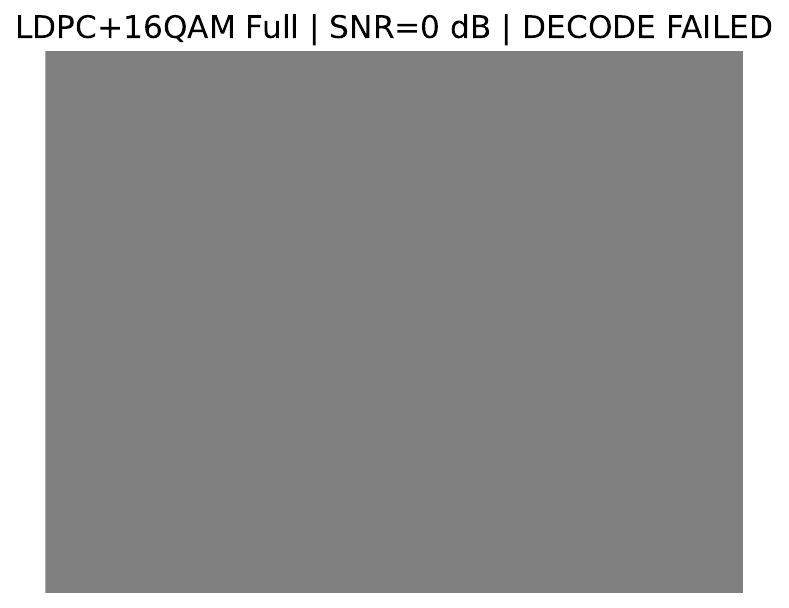}}
        \subfigure[]{\includegraphics[width=0.23\linewidth]{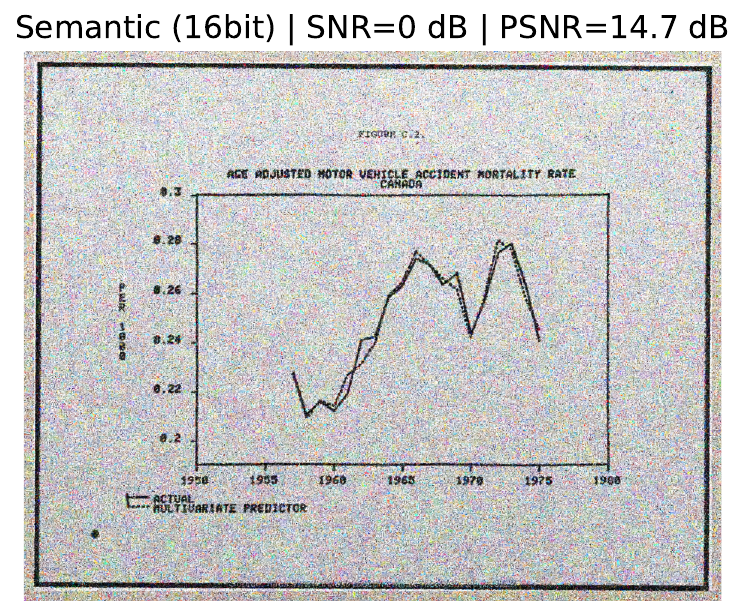}}
        \subfigure[]{\includegraphics[width=0.23\linewidth]{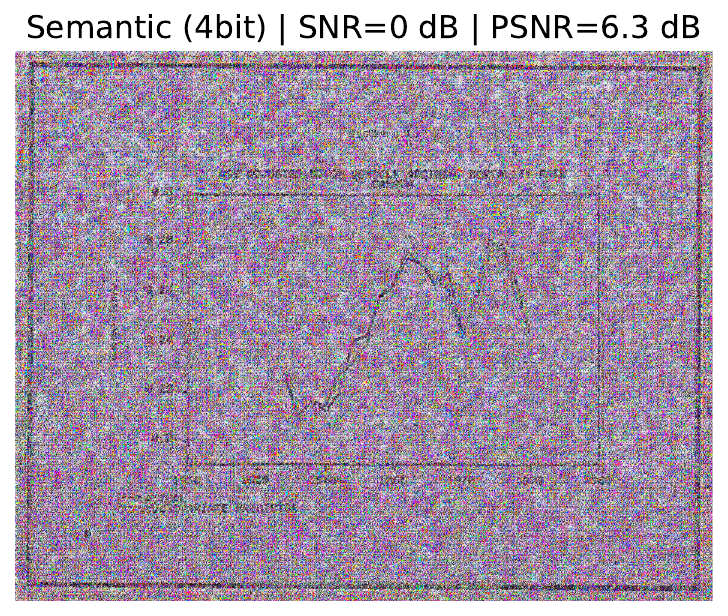}}\\
        \subfigure[]{\includegraphics[width=0.23\linewidth]{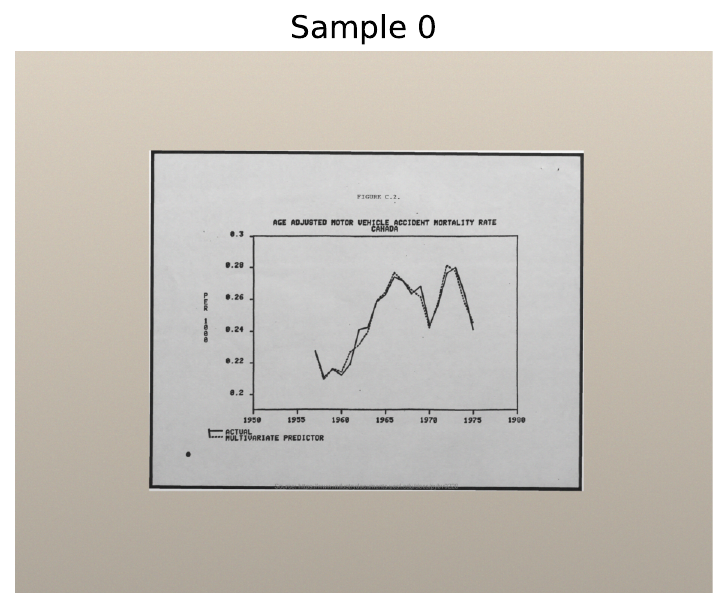}}
        \subfigure[]{\includegraphics[width=0.245\linewidth]{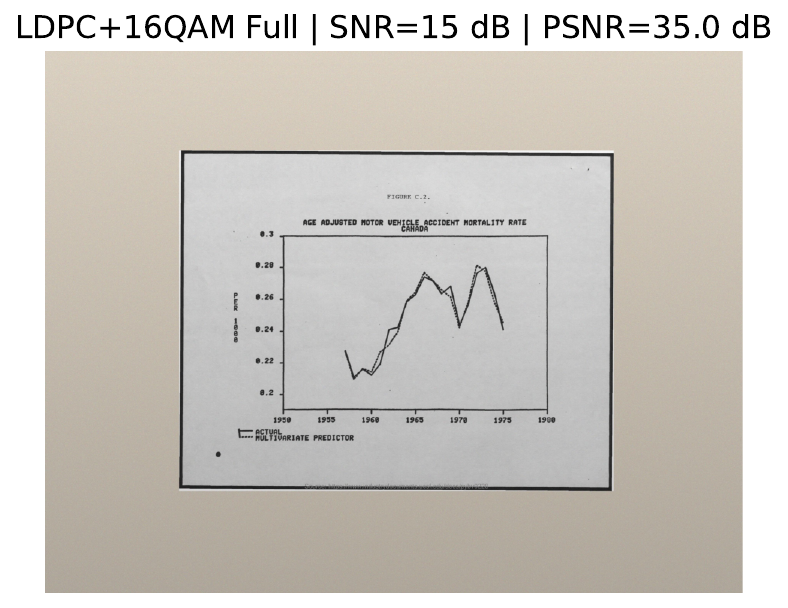}}
        \subfigure[]{\includegraphics[width=0.23\linewidth]{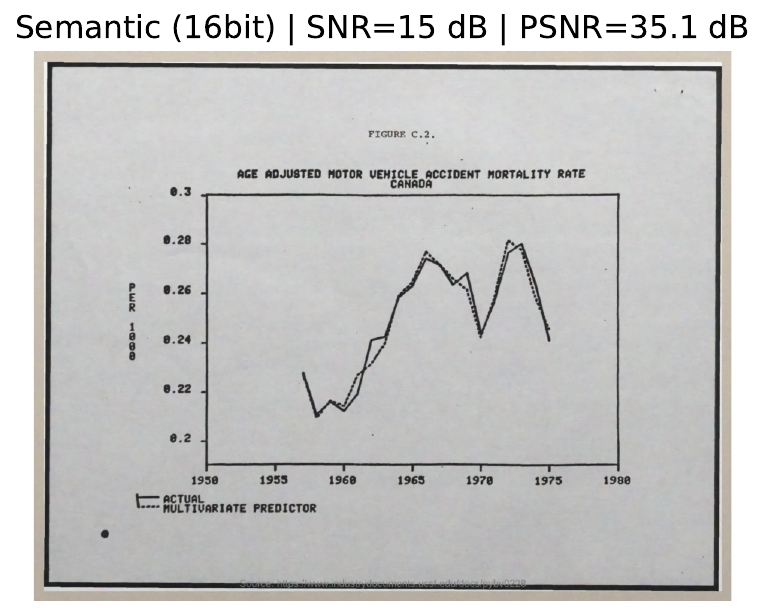}}
        \subfigure[]{\includegraphics[width=0.23\linewidth]{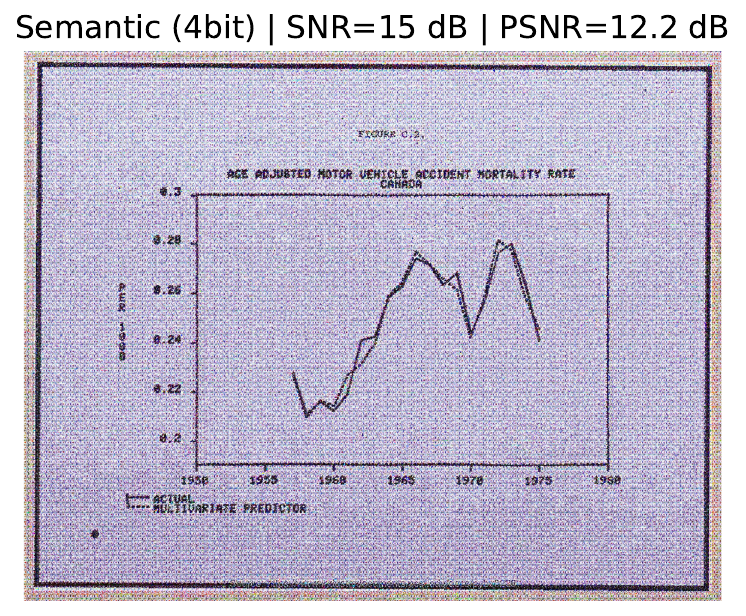}}
	\caption{Examples under different preprocessing or transmission methods, where the semantic methods can still provide task-acceptable visual quality at low SNR.}
	\label{case2_ex}
\end{figure*}
\subsection{Performance of Case II: Document Reading and Answering}

In Case~II, we investigate whether preserving document layout while removing irrelevant background can improve the bandwidth-performance trade-off observed in Case~I. Specifically, the AI glasses apply Canny-based boundary detection to localize the dominant document region, and only the cropped part is transmitted. We compare conventional  and semantic transmission methods under varying compression levels.

Fig. \ref{case2}(a) compares the required complex symbols under different transmission schemes. Transmitting the full image with LDPC+16QAM incurs the highest symbol cost, while Canny-based preprocessing significantly reduces the required bandwidth. Semantic transmission further decreases the symbol count as the quantization is reduced from 16-bit to 4-bit, achieving substantial bandwidth savings compared to conventional digital transmission.

Fig. \ref{case2}(b) shows the corresponding task SR versus SNR. For conventional LDPC+16 QAM transmission with preprocessed image, performance exhibits threshold behavior, with  about zero SR at low SNR and a sharp increase once the SNR exceeds the decoding threshold. In contrast, semantic transmission degrades more gracefully and maintains non-zero SRs even at low SNR. At moderate SNR (5~dB), 16-bit semantic transmission outperforms LDPC with preprocessing, indicating improved robustness at the low SNR. However, aggressive compression  reduces semantic fidelity, leading to noticeable performance degradation at higher SNR. Overall, these results demonstrate that the preprocessing effectively mitigates the structural information loss observed in Case~I, while semantic transmission provides smoother performance adaptation across channel conditions.

Fig. \ref{case2_ex}  shows representative reconstruction results under different schemes and SNR conditions. At low SNR (Fig. \ref{case2_ex} (b)-(d)), conventional LDPC+16QAM transmission fails to decode, leading to complete information loss. The 16-bit semantic transmission still preserves a recognizable document structure, including the overall layout and chart regions despite visible noise. In contrast, aggressive 4-bit semantic compression results in severe distortion, degrading structural clarity. At high SNR (Fig. \ref{case2_ex} (f)-(h)), LDPC transmission achieves near-lossless reconstruction while 16-bit semantic transmission achieves comparable quality, indicating that there is no performance sacrifice under favorable channel conditions. However, 4-bit semantic transmission remains limited by compression (PSNR = 12.2~dB). 

In general, the above experiments show that preserving document layout via Canny-based preprocessing effectively bridges the gap observed in Case~I, reducing bandwidth while retaining the structural information from the document. Compared with conventional transmission, semantic transmission further improves robustness at low SNR through graceful degradation, while achieving comparable performance at high SNR.

\subsection{Performance of Case III:  Scene Watching}
\begin{figure}
	\centering

		\subfigure[Original (640 $\times$ 480) ]{\includegraphics[width=0.45\linewidth]{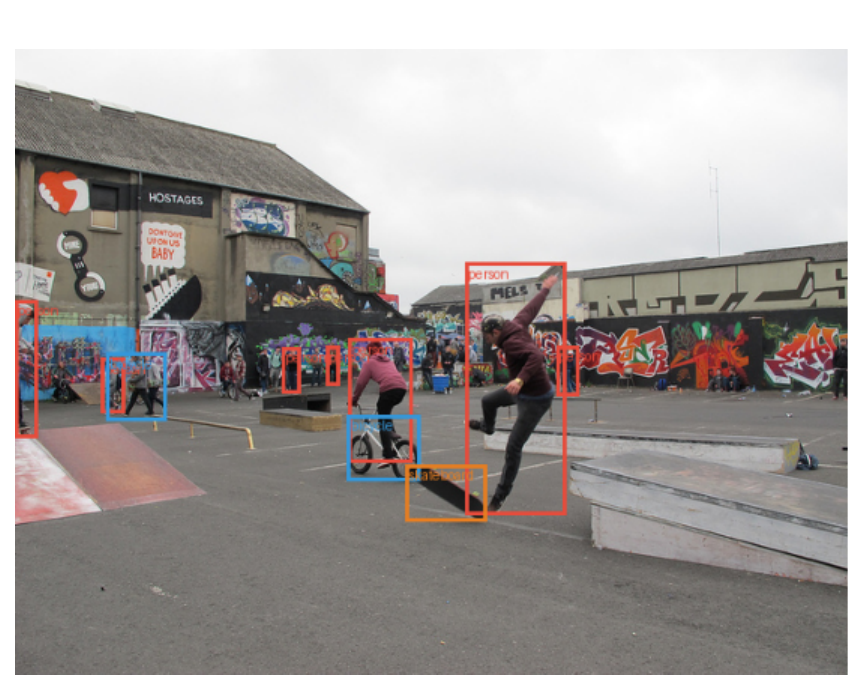}}
        \subfigure[Preprocessed (454 $\times$ 216)]{\includegraphics[width=0.45\linewidth]{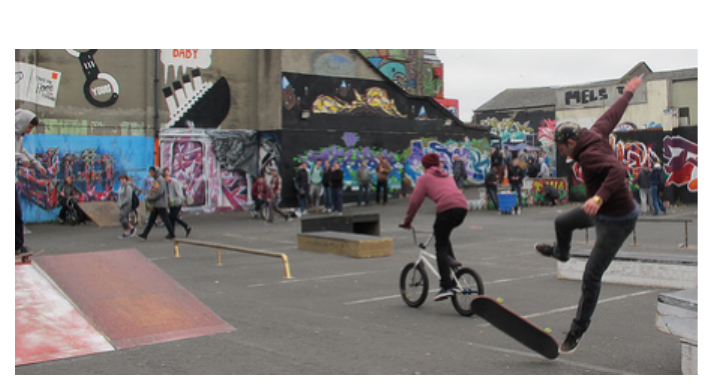}}\\
        \subfigure[Original (640 $\times$ 480)]{\includegraphics[width=0.45\linewidth]{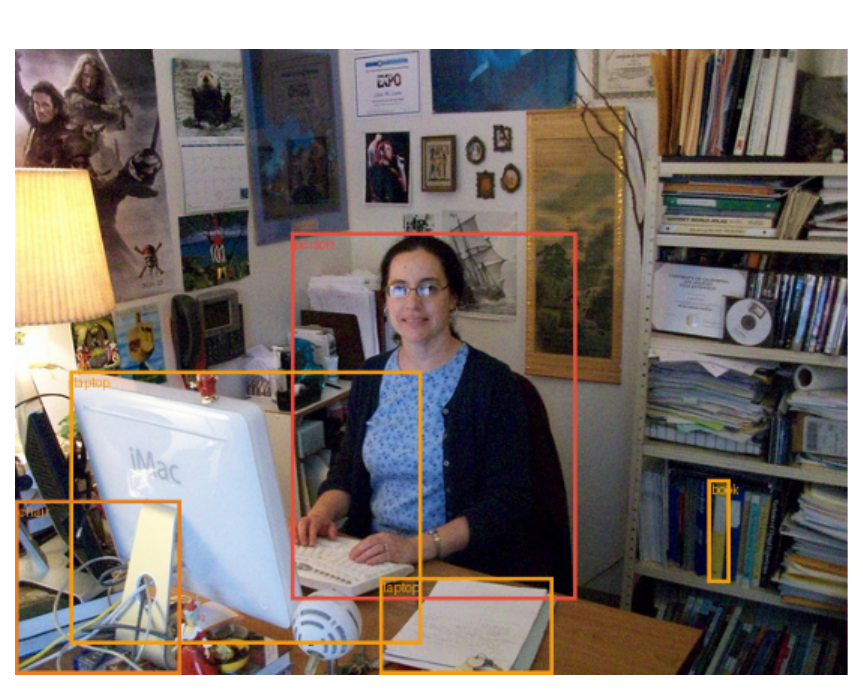}}
        \subfigure[Preprocessed (576 $\times$ 355)]{\includegraphics[width=0.45\linewidth]{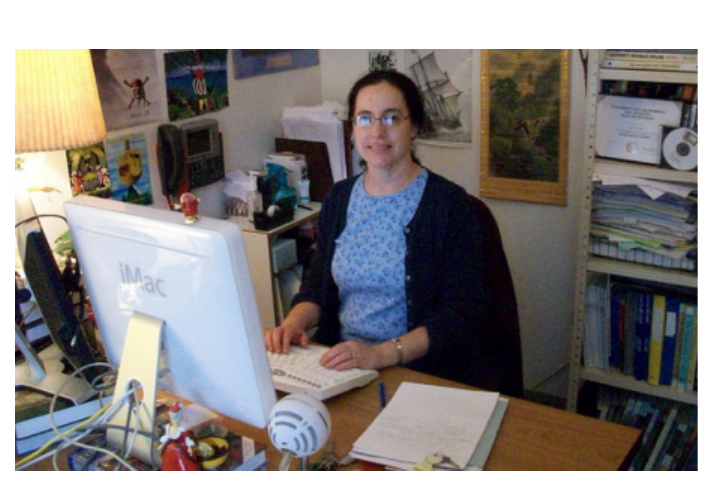}}\\
        \subfigure[Original (480 $\times$ 640)]{\includegraphics[width=0.4\linewidth]{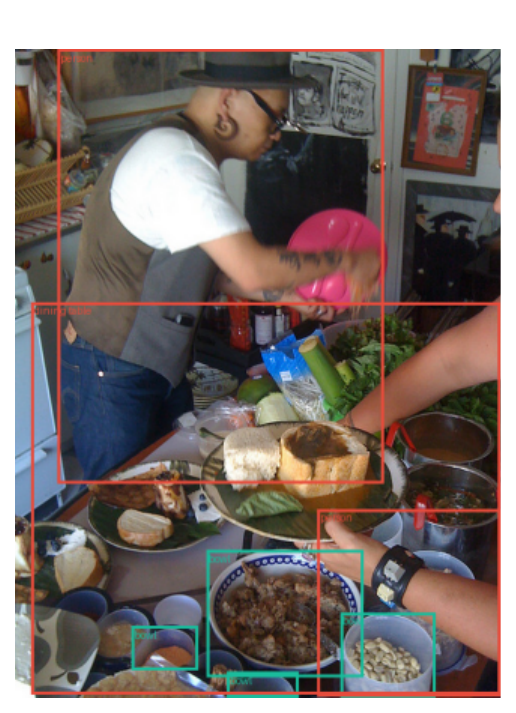}}
        \subfigure[Preprocessed (480 $\times$ 640)]{\includegraphics[width=0.4\linewidth]{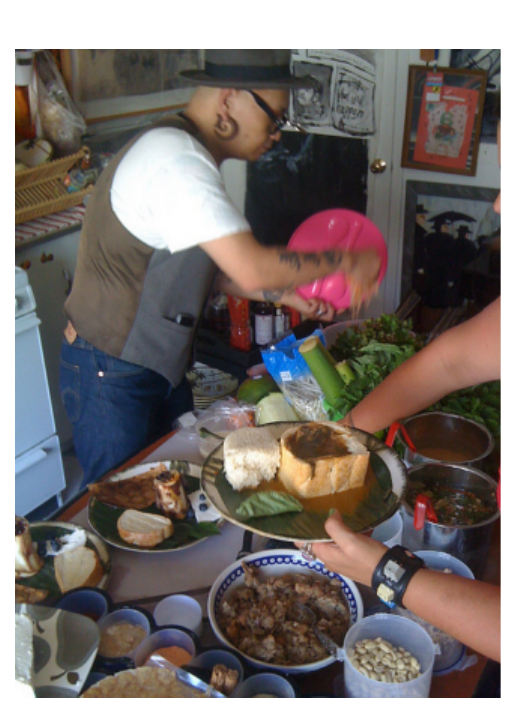}}

	\caption{Examples of Case III, where the main parts of the images are preserved and transmitted to the cloud server. This solution is considered to satisfy the task requirement in most scenarios.}
	\label{case3_ex}
\end{figure}

In Case~III, we extend the study from document understanding to general scene watching. Unlike Cases~I and II, which focus on  documents, this scenario involves natural images with multiple objects and diverse semantic content. The AI glasses employ YOLO-based object detection to localize salient regions, and the merged object-aware crop is transmitted to the server for downstream visual-language reasoning. This setting evaluates whether semantic transmission can preserve high-level scene semantics under bandwidth and channel constraints.

Fig. \ref{case3_ex} illustrates representative YOLO-guided preprocessing results for three scene images. In each case, the detector identifies objects and merges the corresponding bounding boxes into a unified region of interest. For example, in the street scene, the preprocessing step reduces the resolution from \(640 \times 480\) to \(454 \times 216\), eliminating large sky and ground areas while retaining the human activities and interaction regions that dominate scene semantics. In the indoor office, preprocessing suppresses surrounding clutter and focuses on the person and workspace, reducing spatial redundancy without discarding key contextual cues. In the cooking scene, most object regions are densely distributed across the frame. The preprocessed image maintains nearly the original resolution, indicating that the algorithm adaptively preserves full-frame information when semantic content is spatially widespread. These examples demonstrate that YOLO-guided preprocessing achieves adaptive spatial compression and minimal information loss when semantic objects occupy most of the scene.
\begin{figure}[!h]
	\centering

		\subfigure[]{\includegraphics[width=0.99\linewidth]{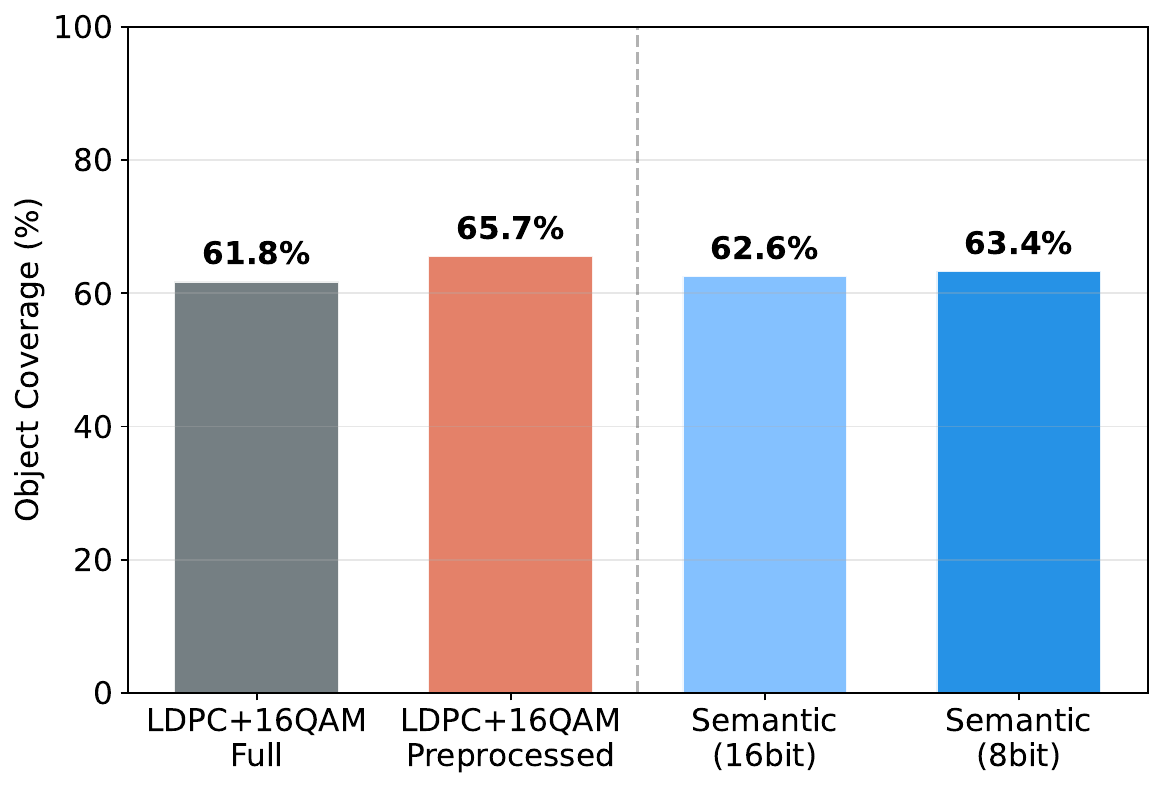}}\\
        \subfigure[]{\includegraphics[width=0.99\linewidth]{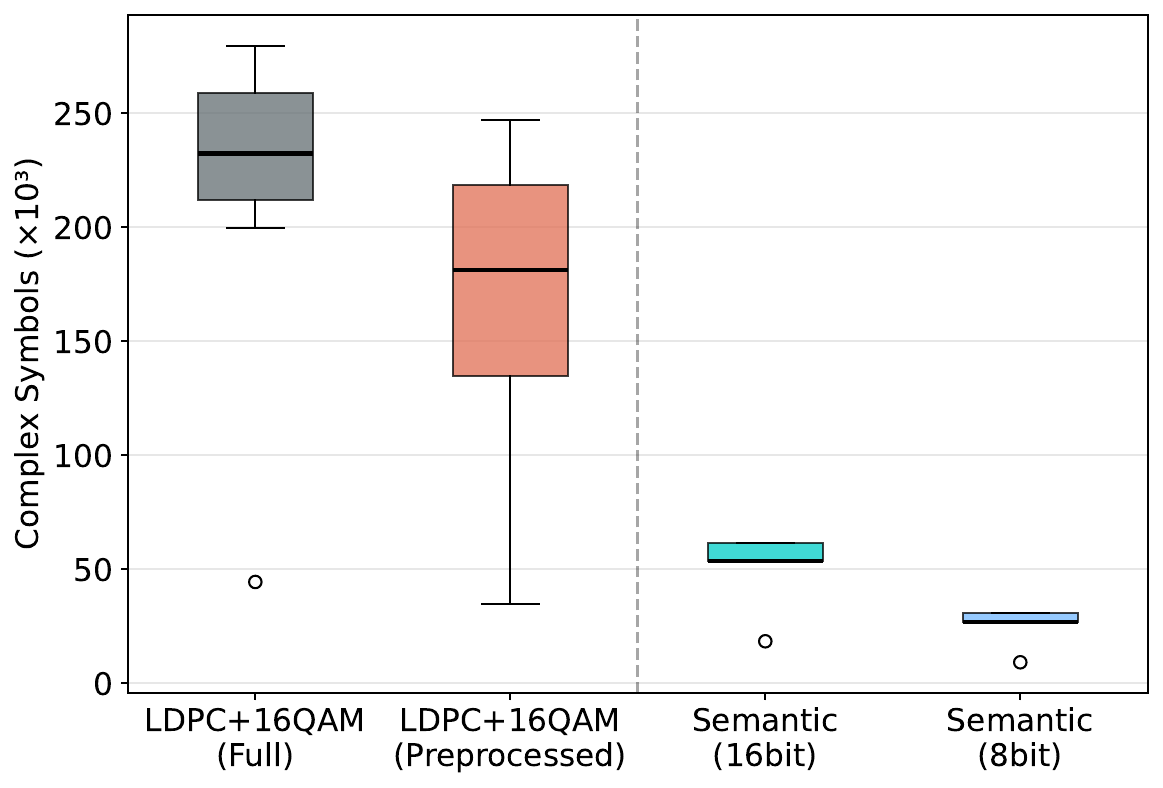}}\\
	\caption{Task performance and bandwidth cost of the proposed methods in Case III: (a) Object coverage of the different preprocessing or transmission methods. (b) Complex symbol cost of different methods. }
	\label{case3}
\end{figure}

Fig.~\ref{case3}(a) compares the object coverage ratio under different transmission schemes. Applying  preprocessing increases the coverage from 61.8\% (LDPC+16QAM full transmission) to 65.7\%, indicating that background removal allows a larger proportion of transmitted pixels to focus on semantic objects. The semantic schemes (16-bit and 8-bit) achieve comparable coverage levels, demonstrating that semantic compression does not distort object-aware spatial allocation. Fig.~\ref{case3}(b) further shows the required complex symbols. When the LDPC transmission incurs the highest symbol cost, preprocessing still reduces the transmission load. Semantic transmission provides a substantially larger reduction, with 16-bit and 8-bit schemes requiring only a small fraction of the symbols needed for conventional digital transmission. Notably, the reduction in symbol cost does not correspond to a decrease in object coverage, indicating that YOLO-guided preprocessing effectively improves semantic efficiency by concentrating transmission resources on object-relevant regions. Overall, Fig.~\ref{case3} confirms that intention-aware preprocessing enhances semantic density while enabling substantial bandwidth savings.

Table~\ref{tab2} reports the BERTScore under different SNR levels for conventional LDPC transmission and semantic transmission with 16-bit and 8-bit representations. At low SNR (0~dB), semantic transmission achieves higher BERTScore than LDPC, indicating stronger robustness when channel conditions are poor. At moderate SNR (5~dB), semantic 16-bit attains the best performance, while LDPC remains lower. As SNR increases to 10~dB and 15~dB, the performance gap narrows, and all schemes converge to similar BERTScore values around 0.91. That means  both transmission paradigms preserve sufficient semantic information  under reliable channel conditions. Notably, the 8-bit semantic scheme maintains competitive performance across all SNR levels, demonstrating that moderate semantic compression can substantially reduce transmission cost without degrading high-level scene understanding.

\begin{table}[htbp]
\centering

\caption{BERTScore under Different SNR Levels}
\begin{tabular}{c|rrrr}
\toprule
SNR& \textbf{0 dB} & \textbf{5 dB} & \textbf{10 dB} & \textbf{15 dB} \\
\midrule
\makecell{LDPC+16QAM \\ (Preprocessed)} & 0.8820 & 0.8824 & 0.9111 & 0.9094 \\ \midrule
\makecell{Semantic \\ (16 bit)} & 0.9131 & \textbf{0.9148} & 0.9117 & 0.9079 \\ \midrule
\makecell{Semantic \\ (8 bit)} & \textbf{0.9109} & 0.9126 & \textbf{0.9130} & \textbf{0.9142} \\
\bottomrule
\end{tabular}
\label{tab2}
\end{table}

Overall, Case~III demonstrates that YOLO-based intention-aware preprocessing improves semantic transmission efficiency in complex scenes. By concentrating on object-relevant regions, the system increases effective semantic density while reducing redundant background transmission. Combined with semantic encoding, this approach achieves substantial bandwidth savings without compromising high-level scene understanding.

\subsection{Ablation Study of Task Activation Modes}

This subsection presents a simple example to compare different commands for the AI glasses, where the proposed intention-aware pipeline without the stored commands and direct user voice commands are compared. The scene watching scenario is also tested in this part.

The intention-aware method relies only on the VLM to analyze the current image and infer the user's potential task, followed by selecting corresponding preprocessing tools. To simulate historical interactions, we additionally introduce stored commands generated by randomly sampling realistic VLM task instructions (e.g., scene description, text translation, object counting). These commands follow typical user voice-query formats and are maintained in a lightweight memory buffer. During inference, the most semantically relevant stored command is retrieved based on embedding similarity and injected into the pipeline as auxiliary prompt guidance. Finally, the user can directly specify the task through voice input (ground-truth instruction). Obviously, a higher degree of alignment between the perceived intention and the actual user instruction leads to improved processing efficiency within the transmission system.

\begin{table}[htbp]
\centering

\caption{Different commands for the AI glasses}
\begin{tabular}{c|rr}

\toprule
Different commands& SR& Bandwidth   \\
\midrule
\textbf{Full Image} & 95.0\% & 50.7 KBytes  \\ \midrule
\textbf{Intention-Aware} & 86.7\% & 38.4 KBytes  \\ \midrule
\textbf{Intention-Aware(Stored)} & \textbf{93.3\%} & 24.0 KBytes   \\ \midrule
\textbf{Direct Voice} & \textbf{93.3\%} & \textbf{8.6 KBytes}   \\ 
\bottomrule
\end{tabular}
\label{tab3}
\end{table}

\begin{figure*}[h]
	\centering
	{\includegraphics[width=0.8\linewidth]{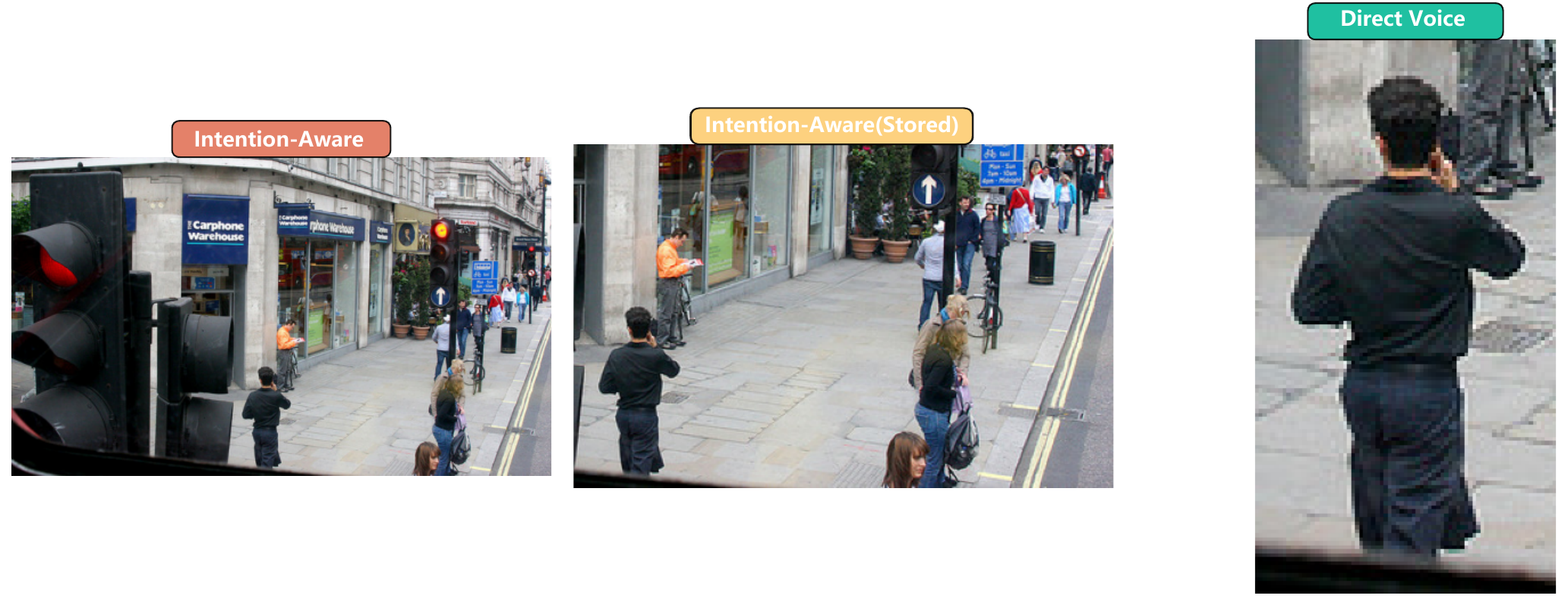}}

	\caption{Examples of intention-aware preprocessed image and the competing ones.}
	\label{EX_4}
\end{figure*}

As shown in Table~\ref{tab3},  ``Full Image'' denotes the conventional baseline where the entire image is compressed using JPEG with the optimized quality and transmitted with LDPC channel coding under the same fixed 10 dB SNR as other methods. The proposed intention-aware transmission with stored commands achieves a 93.3\% SR while requiring only 24.0 KBytes, significantly reducing bandwidth compared with full image transmission (50.7 KBytes). 
When the intention-aware module operates without stored commands, the SR drops to 86.7\%, indicating that historical commands  play a critical role in stabilizing intention inference. In contrast, direct command transmission achieves the same SR  with the lowest bandwidth, since explicit user input eliminates intention ambiguity and allows fully focused semantic extraction.  However, the detailed command will be analyzed by the VLM and cannot be stored if it may lead to the intention mistake in the future. This design can be explianed with the following example.

As shown in Fig. \ref{EX_4}, when only the current visual input is available, the intention-aware pipeline infers that ``the user intends to observe the pedestrian crossing'', corresponding to the most salient semantic region in the scene, which results in a relatively coarse intention estimation based solely on visual cues. When historical user commands, such as ``Observe pedestrians'' are incorporated, prior interaction context can be leveraged to refine the inferred intention. This enables more precise identification of task-relevant regions and further suppression of irrelevant background content during compression. Although direct voice input ``Observe the person in black clothing'' provides explicit and accurate task descriptions that facilitate reliable intention estimation, continuous user interaction may not always be feasible in practical deployments. Therefore, utilizing historical command information offers a practical and scalable means to enhance intention awareness while reducing reliance on real-time user input. Importantly, ``Observe pedestrians'' is considered to be stored while the ``Observe the person in black clothing'' is not a good choice because the detailed command is not robust for the other tasks.

Overall, the results highlight the trade-off between interaction overhead and transmission efficiency. While direct voice commands achieve the lowest bandwidth, the intention-aware pipeline with stored commands attains comparable accuracy with reduced user involvement, thereby validating the benefit of the intention-aware pipeline for efficient wearable semantic communication.

\section{Conclusion}

In this paper, we proposed an intention-aware semantic communication framework for AI glasses systems under an edge-cloud collaborative architecture. We introduced an intention-aware pipeline and developed an agent-oriented wearable-cloud architecture that reformulates wireless communication as a semantic interface between an edge semantic agent and a cloud-based cognitive agent. To support efficient uplink transmission under  wireless constraints, we designed an intention-aware semantic image transmission mechanism that dynamically adjusts spatial resolution and semantic granularity according to inferred intention. By selectively transmitting semantic representations instead of full-resolution images, the proposed framework achieves bandwidth-efficient communication while preserving downstream inference performance. Our results show that intention-based preprocessing reduces uplink bandwidth  by 50\% without  significant performance loss, demonstrating the practical viability of intention-aware wearable AI communication. In the future, multi-user communications in wearable AI systems with the large task space or the sufficient preprocessing tools will be investigated.

	\bibliographystyle{IEEEtran}
	\bibliography{bibtex0320}
	
	%
	
	
	
\end{document}